%% file: main.tex
\documentclass{article}

\usepackage{microtype}
\usepackage{graphicx}
\usepackage{subfig}
\usepackage{booktabs} %
\usepackage{enumitem}

\usepackage{hyperref}

\usepackage[accepted]{icml2021}

\usepackage[utf8]{inputenc}
\DeclareRobustCommand*{\IEEEauthorrefmark}[1]{\raisebox{0pt}[0pt][0pt]{\textsuperscript{\footnotesize\ensuremath{\ifcase#1\or *\or \dagger\or \ddagger\or%
    \mathsection\or \mathparagraph\or \|\or **\or \dagger\dagger%
    \or \ddagger\ddagger \else\textsuperscript{\expandafter\romannumeral#1}\fi}}}}

\usepackage{float}
\usepackage{multirow}
\usepackage{rotating}

\hypersetup{
    colorlinks=true,
    linkcolor=blue,
    filecolor=magenta,      
    urlcolor=cyan,
}
\usepackage[hang,flushmargin]{footmisc}
\usepackage{amsmath}
\usepackage{etoolbox}
\usepackage{booktabs}

\DeclareMathOperator*{\argmax}{arg max}

\newcommand{\sectionscenter}{\let\section\centeredsection}%
\makeatletter
\patchcmd{\@makecaption}
  {\scshape}
  {}
  {}
  {}
\makeatother
\usepackage[]{subfig}
\usepackage{amssymb}
\usepackage{mathtools}
\usepackage{tabularx}
\makeatletter
\def\footnoterule{\kern-3\p@
  \hrule \@width 2in \kern 2.6\p@} %
\makeatother
\makeatletter
\usepackage{textcomp}

\input{macros.tex}

\DeclareMathAlphabet{\mathcal}{OMS}{cmsy}{m}{n}
\icmltitlerunning{Label-Only Membership Inference Attacks}

\begin{document}

\twocolumn[
\icmltitle{Label-Only Membership Inference Attacks}

\begin{icmlauthorlist}
\icmlauthor{Christopher A. Choquette-Choo}{to}
\icmlauthor{Florian Tramèr}{goo}
\icmlauthor{Nicholas Carlini}{ed}
\icmlauthor{Nicolas Papernot}{to}
\end{icmlauthorlist}

\icmlaffiliation{to}{University of Toronto and Vector Institute}
\icmlaffiliation{goo}{Stanford University}
\icmlaffiliation{ed}{Google}

\icmlcorrespondingauthor{Christopher A. Choquette-Choo}{choquette.christopher@gmail.com}

\icmlkeywords{Machine Learning, ICML}

\vskip 0.3in
]

\printAffiliationsAndNotice{}  %

\setcounter{page}{1}
\setcounter{section}{0}
\input{content/0_abstract.tex}
\input{content/1_intro.tex}
\input{content/2_problem_definition}

\input{content/3_approach}

\input{content/4_attack_performance}

\input{content/5_new_defenses}

\input{content/6_outlier_analysis}

\input{content/7_discussion_conclusion}

\section*{Acknowledgments}

We thank the reviewers for their insightful feedback. This work was supported by CIFAR (through a Canada CIFAR AI Chair), by NSERC (under the Discovery Program, NFRF Exploration program, and COHESA strategic research network), and by gifts from Intel and Microsoft. We also thank the Vector Institute's sponsors.

\clearpage
{
\footnotesize
\bibliography{main}
\bibliographystyle{icml2021}
}
\clearpage
\appendix
\input{content/8_appendix}
\clearpage
\end{document}

%% file: macros.tex
\usepackage{xspace}
\newcommand{\ie}{i.e.,\@\xspace}

\newcommand{\etal}{et al.\@\xspace}

\newcommand{\augment}{\mathrm{augment}}
\newcommand{\dist}{\mathrm{dist}}
\newcommand{\mi}{MI\@\xspace}
\usepackage{subfig}
\newcommand{\setmuskip}[2]{#1=#2\relax}

\makeatletter
\newcommand\thefontsize[1]{{#1 The current font size is: \f@size pt\par}}
\makeatother

\usepackage{atbegshi,etoolbox}
\newcommand{\discardpages}[1]{%
  \xdef\discard@pages{#1}%
  \AtBeginShipout{%
    \renewcommand*{\do}[1]{%
      \ifnum\value{page}=##1\relax%
        \AtBeginShipoutDiscard%
        \gdef\do####1{}%
      \fi%
    }%
    \expandafter\docsvlist\expandafter{\discard@pages}%
  }%
}

%% file: content/0_abstract.tex
\begin{abstract}

Membership inference is one of the simplest privacy threats faced by machine learning models that are trained on private sensitive data. 
In this attack, an adversary infers whether a particular point was used to train the model, or not, by observing the model's predictions. 
Whereas current attack methods all require access to the model's predicted confidence score, we introduce a label-only attack that instead evaluates the robustness of the model's predicted (hard) labels under perturbations of the input, to infer membership.
Our label-only attack is not only as-effective as attacks requiring access to confidence scores, it also demonstrates that a class of defenses against membership inference, which we call ``confidence masking'' because they obfuscate the confidence scores to thwart attacks, are insufficient to prevent the leakage of private information.
Our experiments show that training with differential privacy or strong $\ell_2$ regularization are the only current defenses that meaningfully decrease leakage of private information, even for points that are outliers of the training distribution.

\end{abstract}

%% file: content/1_intro.tex
\section{Introduction}\label{sec:intro}
Machine learning algorithms are often trained on sensitive or private user information, e.g., medical records~\citep{medicalclassification}, conversations~\citep{bert}, or financial information~\citep{fraud}.
Trained models can inadvertently leak information about their training data~\citep{cornellMI, secretsharer}---violating users' privacy.

In perhaps the simplest form of information leakage,
\emph{membership inference (\mi)}~\citep{cornellMI}
attacks enable an adversary to determine whether or not a data point was used in the training data. 
Revealing just this information can cause harm---it leaks information about specific individuals instead of the entire population. 
Consider
a model trained to learn the link between a cancer patient's morphological data and their reaction to some drug. An adversary 
with
a victim's morphological data and
query access to the trained model cannot directly infer 
if
the victim has cancer. However, inferring that the victim's data was part of the model's training set reveals that the victim indeed has cancer.

Existing \mi attacks exploit the higher
confidence that models exhibit on their training data~\cite{locationmi, demystify, gan, salem2018mlleaks}. 
An adversary queries the model on a candidate data point to obtain the model's confidence and infers the candidate's membership in the training set based on a decision rule. The difference in prediction confidence is largely attributed to \emph{overfitting}~\cite{cornellMI, privacyoverfittingmi}.

A large body of work has been devoted to understanding and mitigating \mi leakage in ML models. Existing defense strategies fall into two broad categories and either
\begin{enumerate}
\vspace{-0.5em}
    \item[(1)] reduce overfitting~\cite{demystify, cornellMI, salem2018mlleaks}; or,
    \vspace{-0.5em}
    \item[(2)] perturb a model's predictions %
    so as to minimize the success of known membership attacks~\cite{adversarialregularization, memguard, yang2020predictionpurification}.
\end{enumerate}
\vspace{-0.5em}
Defenses in (1) use regularization techniques or increase the amount of training data to reduce overfitting. In contrast, the adversary-aware defenses of (2) explicitly aim to minimize the \mi advantage of a particular attack. They do so either by modifying the training procedure (e.g., an additional loss penalty) or the inference procedure after training. These defenses implicitly or explicitly rely on a strategy that we call \emph{confidence-masking}\footnote{Similar to \emph{gradient masking} from the adversarial examples literature~\cite{blackboxml}.},
where the \mi signal in
the model's confidence scores is masked to thwart existing attacks. 

We introduce \emph{label-only \mi attacks}. Our attacks are more general:
an
adversary 
need only
obtain (hard) labels---without prediction confidences---of
the trained model. This threat model is more realistic, 
as ML models deployed in user-facing products need not expose raw confidence scores. Thus, our attacks can be mounted on any ML classifier.

In the label-only setting, a naive baseline 
predicts misclassified points as non-members.
Our focus is 
surpassing this baseline.
To this end, we
will 
have to make \emph{multiple} queries to the target model. We show how to extract fine-grained \mi signal by analyzing a model's robustness to perturbations of the target data, which reveals signatures of 
its
decision boundary geometry.
Our adversary queries the model for predicted labels on augmentations of data points (e.g., 
translations in vision domains) as well as adversarial examples. 

\paragraph{We make the following contributions. } In \S~\ref{ssec:attackperformance}, we introduce the first label-only attacks, which match confidence-vector attacks.  By combining them, we outperform all others.
    In \S~\ref{ssec:breaking-confidence-masking},~\ref{ssec:breaking-memguard} and~\ref{ssec:breakingadversarialregularization}, we show that \emph{confidence masking} is not a viable defense to privacy leakage, by breaking two canonical defenses that use it---MemGuard and Adversarial Regularization. In \S~\ref{sec:generalization}, we evaluate two additional techniques to reducing overfitting: data augmentation and transfer learning. We find that data augmentation can worsen \mi leakage while transfer learning can mitigate it. In \S~\ref{sec:outlier}, we introduce ``outlier \mi'': a stronger property that defenses should
    satisfy to protect \mi of worst-case inputs; at present, differentially private training and (strong) L2 regularization 
    appear to be 
    the only effective defenses. Our code is available at \url{https://github.com/cchoquette/membership-inference}.

%% file: content/2_problem_definition.tex
\section{Background and Related Works}
Membership inference attacks~\cite{cornellMI} are a form of privacy leakage that identify if a given data sample was in a machine learning model's training dataset. 
Given a sample $x$ and access to a trained model $h$, the adversary uses a classifier or decision rule $f_h$ to compute a membership \emph{prediction} $f(x; h) \in \{0, 1\}$, with the goal that $f(x; h) = 1$ whenever $x$ is a training point. The main challenge in mounting a \mi attack is creating the attack classifier $f$, under various assumptions about the adversary's knowledge of $h$ and its training data distribution. Most prior work assumes that an adversary has only black-box access to the trained model $h$, via a query interface that on input $x$ returns part or all of the confidence vector $h(x) \in [0, 1]^C$ (for a classification task with $C$ classes). 

The attack classifier $f$ is often trained on a local shadow (or, source) model $\hat{h}_i$, which is trained on the same (or a similar) distribution as $h$'s training data. Because the adversary trained $\hat{h}_i$, they can assign membership labels to any input $x$, and use this dataset to train $f$. \citet{salem2018mlleaks} later showed that this strategy succeeds even when the adversary only has data from a different, but similar, task
and that shadow models are unnecessary: a threshold predicting $f(x; h) = 1$ 
when
the max prediction confidence,
$\max_i h(x)$,
is above a tuned threshold, suffices.

\citet{privacyoverfittingmi} investigate how querying related inputs $x'$ to $x$ can improve \mi. \citet{song2019privacy} explore how models explicitly trained to be robust to adversarial examples can become more vulnerable to \mi (similar to our analysis of data augmentation in \S~\ref{sec:generalization}). Both works are crucially different because they use a different attack methodology and assume access to the confidence scores. 
\citet{sablayrolles2019white} demonstrate that black-box attacks (like ours) can approximate white-box attacks by effectively estimating the model loss for a data point.
Refer to Appendix \S~\ref{sec:problemdefinition} for a detailed background, including on defenses.

%% file: content/3_approach.tex
\section{Attack Model Design}\label{sec:approach}
 Our proposed \mi attacks improve on existing attacks by (1) combining \emph{multiple 
 strategically perturbed samples (queries)} as a fine-grained signal of
 the model's decision boundary, and (2) operating in a label-only regime. Thus, our attacks pose a threat to \emph{any} query-able ML service.

\subsection{A Naive Baseline: The Gap Attack}
\label{ssec:baseline}
Label-only \mi attacks face a challenge of granularity.
For any query $x$, our attack model's information is limited to only the predicted class-label, $\argmax_i h(x)_i$.
A simple \emph{baseline} attack~\cite{privacyoverfittingmi}---that predicts any misclassified data point as a non-member of the training set---is a useful benchmark to assess the extra (non-trivial) information that \mi
attacks, label-only or otherwise,
 can extract. We call this baseline the \emph{gap attack} because its accuracy is directly related to the gap between the model's accuracy on training data ($\text{acc}_{\text{train}}$) and held out data ($\text{acc}_{\text{test}}$):
 \begin{equation}
 \label{eq:gap}
 1/2 + (\text{acc}_{\text{train}}-\text{acc}_{\text{test}})/2 \;,
 \end{equation}
 where $\text{acc}_{\text{train}}, \text{acc}_{\text{test}} \in [0, 1]$.
To 
exploit additional leakage on top of
this baseline attack (achieve non-trival \mi), any label-only adversary 
must
necessarily 
make additional queries to the model. To the best of our knowledge, this trivial baseline is the only attack proposed 
in prior work that uses only the predicted label, $y = \argmax_i h(x)_i$.

\subsection{Attack Intuition}\label{ssec:attack-intuition}
Our strategy is to compute label-only ``proxies'' for the model's confidence by evaluating its \emph{robustness} to strategic input perturbations of $x$, either synthetic (\ie data augmentation) or adversarial (examples)~\cite{szegedy2013intriguing}.
Following a max-margin perspective, we predict that data points that exhibit high robustness are training data points. Works in the adversarial example literature share a similar perspective that non-training points are closer to the decision boundary and thus more susceptible to perturbations~\cite{tanay2016boundary,tian2018imagetransformation,hu2019new}.

Our intuition for leveraging robustness is two-fold. First, models \emph{trained with} data augmentation have the capacity to overfit to them~\cite{zhang2016understanding}. Thus, we evaluate any ``effective'' train-test gap on the augmented 
dataset by evaluating $x$ and its augmentations,
giving us a more fine-grained \mi signal. For models not trained using augmentation, their robustness to perturbations can be a proxy for model confidence. Given the special case of (binary) logistic regression models, with a learned weight vector ${w}$ and bias $b$, the model will output a confidence score for the positive class of the form: 
$h(x) \coloneqq \sigma(w^\top x + b)$,
where $\sigma(t)=\frac{1}{1+e^{-t}} \in (0,1)$ is the logistic function.

Here, there is a monotone relationship between the confidence at
$x$ and the Euclidean \emph{distance} 
to the model's decision boundary. 
This distance
is $(w^\top x + b)/||w||_2 = \sigma^{-1}(h(x))/||w||_2$.
Thus, obtaining a point's distance to the boundary yields the same information as the 
confidence score. 
Computing this distance 
is exactly the problem of finding the smallest \emph{adversarial perturbation}, which can be done using label-only access to a classifier~\cite{brendel2017decision, chen2019hopskipjumpattack}. 
Our thesis is that this relationship
will persist for deep, non-linear models.
This thesis is supported by prior work that suggests that deep neural networks can be closely approximated by linear functions in the vicinity of the data
~\cite{goodfellowadversairalexamples}.

\subsection{Data Augmentation}
\label{ssec:augmentation_attack_design}
Our data augmentation attacks create a \mi classifier $f(x;h)$ for a model $h$. Given a target point
$(x_0, y_{\text{true}})$, the adversary trains $f$ to output $f(x_0,h)=1$, if $x_0$ was a training member. To do this, they tune $f$ to maximize \mi accuracy on a source (or "shadow") model assuming knowledge of the target model's architecture and training data distribution. 
They then 
``transfer'' $f$ to 
perform \mi by querying the black-box model $h$.
Using $x_0$,
we create additional data points $\{\hat{x}_1, \dots, \hat{x}_N\}$ via different data augmentation strategies, described 
below.
We 
query the target model $h$ ($\hat{h}$ in tuning) 
to obtain labels $(y_0, y_1, \dots, y_N) \gets (h(x), h(\hat{x}_1), \dots h(\hat{x}_N))$. Let $b_i \gets 
\mathbb{I}\left(y_{true} = (y_i)\right)$
be the indicator function for whether the i-th queried point was misclassified. 
Finally, we apply $f(b_0, \dots, b_N) \to \{0,1\}$ to 
classify $x_0$. 

We experiment with two common data augmentations in the computer vision domain: image rotations and translations. For \textbf{rotations}, we generate $N=3$ images as rotations by a magnitude $\pm r^\circ$ for $r\in [1, 15]$. For \textbf{translations}, we generate $N=4d+1$ translated images  satisfying $|i| + |j| = d$ for a  pixel bound $d$, where we horizontal shift by $\pm i$ and vertical shift by $\pm j$.  In both we include the source image.

\subsection{Decision Boundary Distance}
\label{ssec:decisionboundarydistance}

These attacks predict membership using a point's distance to the model's decision boundary. Here, we extend the intuition that this distance can be a proxy for confidence of linear models (see \S~\ref{ssec:attack-intuition}) to deep neural networks.

Recall that confidence-thresholding attacks predict highly confident samples as members~\cite{salem2018mlleaks}. Given some estimate $\dist_h(x, y)$ of a point's $\ell_2$-distance to the model's boundary, we predict $x$ a member 
if $\dist_h(x, y) > \tau$ for some threshold $\tau$. We define $\dist_h(x, y)=0$ for misclassified points, where $\argmax_i h(x)_i \neq y$, because no perturbation was
needed.
We tune $\tau$ on a shadow $\hat{h}_i$, and find that even crude estimates, e.g., Gaussian noise, can lead to nearly comparable attacks (see \S~\ref{ssec:label-only-query}).
We now discuss methods for estimating $\dist(x, y)$.

\paragraph{A White-Box Baseline} for estimating $\dist(x, y)$ is an idealized \emph{white-box} attack
and is therefore not label-only.
We use adversarial-examples generated by the Carlini and Wagner attack~\cite{carliniwagner}: given $(x, y)$ the attack tries to find the closest point $x'$ to $x$ in the Euclidean norm, such that $\argmax h(x') \neq y$.

\paragraph{Label-only attacks} use only black-box access. We rely on label-only adversarial example attacks~\cite{brendel2017decision, chen2019hopskipjumpattack}. These attacks start from a random point $x'$ that is misclassified, i.e., $h(x') \neq y$. They then ``walk'' along the boundary while minimizing the distance to $x$. We use ``HopSkipJump''~\cite{chen2019hopskipjumpattack}, which closely approximates stronger white-box attacks.

\paragraph{Robustness to random noise} is a simpler approach based on random perturbations. Again, our intuition stems from linear models: a point's distance to the boundary is directly related to the model's accuracy when it is perturbed by isotropic Gaussian noise~\cite{ford2019adversarial}. 
We compute a proxy for $d_h(x, y)$ by evaluating the accuracy of $h$ on $N$ points
$\hat{x}_i = x + \mathcal{N}(0, \sigma^2 \cdot I)$, where $\sigma$ is tuned on 
$\hat{h}$.
For binary features we instead use Bernoulli noise: each $x_j \in x$ is flipped with probability $p$, which is tuned on $\hat{h}$.

\paragraph{Many signals for robustness} can be combined to improve the attack performance. 
We evaluate $d_h(x, y)$ for augmentations of $x$ from \S~\ref{ssec:augmentation_attack_design}.
We only evaluate this attack where indicated due to its high query cost (see \S~\ref{ssec:label-only-query}).

%% file: content/4_attack_performance.tex
\section{Evaluation Setup}
\label{sec:evalsetup}
Our evaluation is aimed at understanding how label-only \mi attacks compare with prior attacks that rely on access to a richer query interface. To this end, \textbf{we use an identical evaluation setup as prior works}~\cite{shokriminmax,cornellMI,long2017towards,demystify,salem2018mlleaks} (see Appendix \S~\ref{ssec:ourthreatmodel}). We answer the following questions in our evaluation, \S~\ref{sec:labelonlyeval}, \S~\ref{sec:generalization} and \S~\ref{sec:outlier}:
\begin{enumerate}[topsep=0pt,itemsep=0ex,partopsep=1ex,parsep=1ex]
    \item Can label-only \mi attacks match prior attacks that use the model's (full) confidence vector? 
    \item Are defenses against confidence-based MI attacks always effective against label-only attacks?
    \item What is the query complexity of  label-only attacks?
    \item Which defenses prevent both label-only and full confidence-vector attacks?
\end{enumerate}
\vspace{-0.25em}
To evaluate an attack's success, we pick a balanced set of points from the task distribution, of which half come from the target model's training set.
We measure attack success as overall \mi accuracy but find F1 scores to approximately match, with near 100\% recall. See Supplement \S~\ref{ssec:on-measuring-success} for further discussion on this evaluation.

Overall, we stress that our main goal is to show that \emph{in settings where \mi attacks have been shown to succeed}, a label-only query interface is sufficient. In general, we should not expect our label-only attacks to \emph{exceed} the performance of prior \mi attacks since the former uses strictly less information from queries than the latter. 
There are three notable exceptions: our combined attack\footnote{Note that this attack's performance exceeds prior confidence-vector attacks, but that we do not test its confidence-vector analog. Our results indicate that it should perform comparably.} (\S~\ref{ssec:attackperformance}), ``confidence masking'' defenses (\S~\ref{ssec:breaking-confidence-masking}), and models trained with significant data augmentation (\S~\ref{ssec:augmentations}). In the latter two cases, we find that existing attacks severely underestimate \mi.

\subsection{Attack Setup}
We provide a detailed account of model architectures and training procedures
in Supplement \S~\ref{app:generating-data} and of our threat model in Supplement \S~\ref{app:threatmodel}. We evaluate our attacks on $8$ datasets used by the canonical work of~\citet{cornellMI}. These include $3$ computer vision tasks\footnote{{\fontsize{8}{0}MNIST, CIFAR-10, and CIFAR-100: \url{https://www.tensorflow.org/api_docs/python/tf/keras/datasets}}}, which are our main focus because of the importance of data augmentation to them, and $4$ non-computer-vision tasks\footnote{{\fontsize{8}{0}Adult Dataset: \url{http://archive.ics.uci.edu/ml/datasets/Adult}\newline Texas-100, Purchase-100, and Locations datasets: \url{ https://github.com/privacytrustlab/datasets}}} to showcase our attacks' transferability. 
We train target neural networks on subsets of the original training data, exactly as performed by~\citet{cornellMI} and several later works (in both data amount and train-test gap). Controlling the training set size 
lets us control the amount of overfitting, which strongly influences the strength of \mi attacks~\cite{privacyoverfittingmi}. Prior work has almost exclusively studied (confidence-based) \mi attacks on these small datasets where models exhibit a high degree of overfitting.
Recall that our goal is to show that label-only attacks match confidence-based approaches: scaling \mi attacks (whether confidence-based or label-only) to larger training datasets is an important area of future work.

\vspace{-0.095em}
\section{Evaluation of Label-Only Attacks}\label{sec:labelonlyeval}
\subsection{Label-Only Attacks Match Confidence-Vector Attacks}\label{ssec:attackperformance}
We first focus on question 1).
Recall from \S~\ref{ssec:baseline} that any label-only attack (with knowledge of $y$) is always trivially lower-bounded by the baseline gap attack of~\citet{privacyoverfittingmi}, predicting any misclassified point as a non-member.

Our main result is that our label-only attacks consistently outperform the gap attack and perform on-par with prior confidence-vector attacks; by combining attacks, we can even \textbf{surpass} the canonical confidence-vector attacks.

\begin{figure}[htb]
    \centering
    \includegraphics[width=0.9\columnwidth]{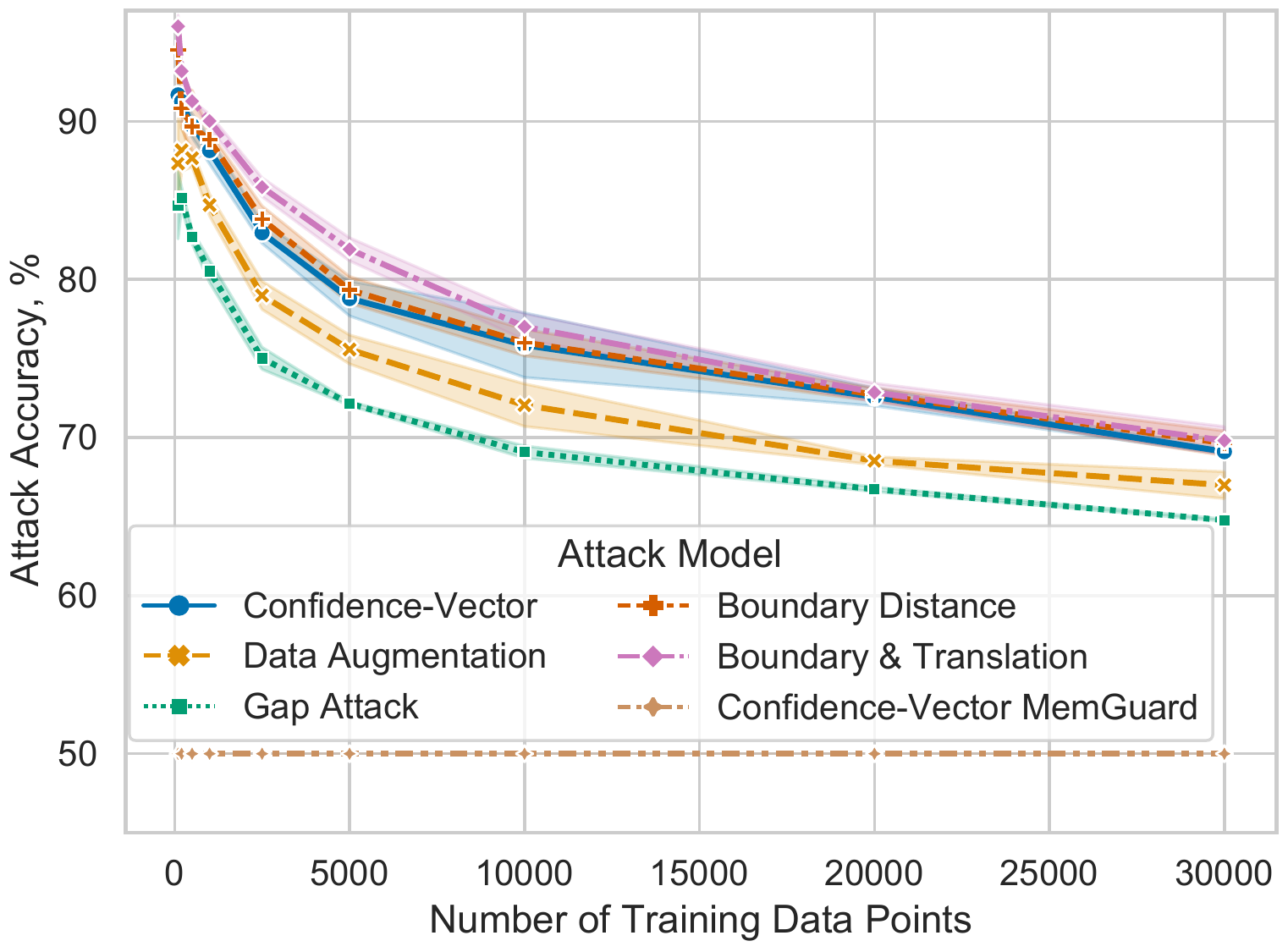}
    \caption{\textbf{Accuracy of \mi attacks on CIFAR-10}. We evaluate $100$ to $10{,}000$ training points and compare the baseline gap attack, the confidence-vector attack that relies on a fine-grained query interface, and our label-only attacks based on data augmentation and distance to the decision boundary. We also show the confidence-vector attack performance against MemGuard, noting that our label-only performances remain nearly unaltered.
    For the data augmentation attack, we report the best accuracy across multiple values of $r$ (rotation angle) and $d$ (number of translated pixels).}
    \label{fig:labelonly}
\end{figure}

Observing Figure~\ref{fig:labelonly} and Table~\ref{tab:cifar100} (a) and (c), we see that the \emph{confidence-vector attack} outperforms the baseline \emph{gap attack}, demonstrating that it exploits non-trivial \mi. Remarkably, we find that our \emph{label-only boundary distance attack} 
performs at least on-par with the confidence-vector attack. Moreover, our simpler but more query efficient (see \S~\ref{ssec:label-only-query}) \emph{data augmentation attacks} also consistently outperform the baseline but fall short of the confidence-vector attacks. Finally, combining these two label-only attacks, we can consistently outperform every other attack.
These models were not trained with data augmentation; in \S~\ref{ssec:augmentations}, we find that when they are, our data augmentation attacks outperform all others. Finally, we verify that as the training set size increases, all attacks monotonically decrease because the train-test gap is reduced.
Note that on CIFAR-100, we experiment with the largest training subset possible: $30{,}000$ data points, since we use the other half as the source model training set (and target model non-members).

\paragraph{Beyond Images}\label{ssec:beyond-images}
We show that our label-only attacks can be applied outside of the image domain in Table~\ref{tab:noiserobustness}. Our label-only attack evaluates a model's accuracy under \emph{random perturbations}, by adding Gaussian noise for continuous-featured inputs, and flipping binary values according to Bernoulli noise (see \S~\ref{ssec:decisionboundarydistance}). Using $10{,}000$ queries, our attacks closely match (at most $4$ percentage-point degradation) confidence-based attacks. Note that our attacks could also be instantiated in audio or natural language domains, using existing adversarial examples attacks~\cite{carlini2018audio} and data augmentations~\cite{nlpaugment}.

\begin{table}[htb]
\vspace{-0.5em}
\caption{\textbf{Accuracy of \mi attacks on CIFAR-100 and MNIST.} The target models are trained using $30{,}000$ data points for CIFAR-100 and $1{,}000$ for MNIST. Tables (a) and (c) report results without any defense; (b) and (d) with MemGuard~\cite{memguard}, which prevents the confidence-vector attacks via ``confidence masking''. `Combined' refers to the boundary and translation attack. Results that are affected by confidence masking are marked in red.}
    \vspace{0.5em}
    \centering
    \small
    \subfloat[CIFAR-100 Undefended]{
    \begin{tabular}{@{} l r @{}}
        Attack & Accuracy \\
        \toprule
        Gap attack & $83.5$\\
        Confidence-vector & ${88.1}$ \\ %
        Data augmentation &  $84.6$ \\ %
        Boundary distance & ${88.0}$\\
        Combined & ${89.2}$
        \end{tabular}}
    \quad\quad
    \subfloat[CIFAR-100 MemGuard]{
    \begin{tabular}{@{} l r @{}}
        Attack & Accuracy \\
        \toprule
        Gap attack & $83.5$\\
        Confidence-vector & {\textcolor{red}{$50.0$}}\\
        Data augmentation &  $84.6$\\
        Boundary distance &  ${88.0}$\\
        Combined & ${89.2}$
    \end{tabular}}
    \\
    \vspace{-6pt}
    \centering
    \subfloat[MNIST Undefended]{
    \begin{tabular}{@{} l r @{}}
        Attack & Accuracy \\
        \toprule
        Gap attack & $53.2$\\
        Confidence-vector & ${55.7}$ \\
        Data augmentation &  $53.9$ \\ %
        Boundary distance & ${57.8}$\\
        Combined & ${58.7}$
    \end{tabular}}
    \quad\quad
    \subfloat[MNIST MemGuard]{
    \begin{tabular}{@{} l r @{}}
        Attack & Accuracy \\
        \toprule
        Gap attack & $53.2$\\
        Confidence-vector & {\textcolor{red}{$50.0$}} \\
        Data augmentation &  $53.9$ \\ %
        Boundary distance & ${57.8}$\\
        Combined & ${58.7}$
    \end{tabular}}
    \label{tab:cifar100}
    \vspace{-1em}
\end{table}

\begin{table}[htb]
    \centering
    \caption{\textbf{Accuracy of membership inference attacks on Texas, Purchase, Location, and Adult.} Where augmentations may not exist, noise robustness can still perform on or near par with confidence-vector attacks. The target models are trained exactly as in~\cite{cornellMI}: $1,600$ points for Location and $10,000$ for the rest. Our noise robustness attack uses $10,000$ queries. Tables (a), (c), (e), and (g) report results without any defense. Tables (b), (d), (f), and (h) report results with MemGuard~\cite{memguard}, which prevents the confidence-vector attacks via ``confidence-masking''. Results that are affected by confidence masking are marked in red.}
    \vspace{0.5em}
    \subfloat[Texas Undefended]{
    \begin{tabular}{@{} l r @{}}
        Attack & Accuracy \\
        \toprule
        Gap attack & $73.9$\\
        Confidence-vector & ${84.0}$ \\ %
        Noise Robustness & ${80.3}$
    \end{tabular}}
    \subfloat[Texas MemGuard]{
    \begin{tabular}{@{} l r @{}}
        Attack & Accuracy \\
        \toprule
        Gap attack & $73.9$\\
        Confidence-vector & {\textcolor{red}{$50.0$}}\\
        Noise Robustness &  ${80.3}$
    \end{tabular}}
    \\
    \centering
    \subfloat[Purchase Undefended]{
    \begin{tabular}{@{} l r @{}}
        Attack & Accuracy \\
        \toprule
        Gap attack & $67.1$\\
        Confidence-vector & ${86.1}$ \\
        Noise Robustness & ${87.4}$ \\
    \end{tabular}}
    \subfloat[Purchase MemGuard]{
    \begin{tabular}{@{} l r @{}}
        Attack & Accuracy \\
        \toprule
        Gap attack & $67.1$\\
        Confidence-vector & {\textcolor{red}{$50.0$}} \\
        Noise Robustness & ${87.4}$ \\
    \end{tabular}}
    \\
    \centering
    \subfloat[Location Undefended]{
    \begin{tabular}{@{} l r @{}}
        Attack & Accuracy \\
        \toprule
        Gap attack & $72.1$\\
        Confidence-vector & ${92.6}$ \\
        Noise robustness &  ${89.2}$ \\ %
    \end{tabular}}
    \subfloat[Location MemGuard]{
    \begin{tabular}{@{} l r @{}}
        Attack & Accuracy \\
        \toprule
        Gap attack & $72.1$\\
        Confidence-vector & {\textcolor{red}{$50.0$}} \\
        Noise Robustness &  ${89.2}$ \\ %
    \end{tabular}}
    \\
    \centering
    \subfloat[Adult Undefended]{
    \begin{tabular}{@{} l r @{}}
        Attack & Accuracy \\
        \toprule
        Gap attack & $58.7$\\
        Confidence-vector & ${59.9}$ \\
        Noise Robustness & ${58.7}$
    \end{tabular}}
    \subfloat[Adult MemGuard]{
    \begin{tabular}{@{} l r @{}}
        Attack & Accuracy \\
        \toprule
        Gap attack & $58.7$\\
        Confidence-vector & {\textcolor{red}{$50.0$}} \\
        Noise Robustness & ${58.7}$
    \end{tabular}}
    \label{tab:noiserobustness}
\end{table}

\subsection{Breaking Confidence Masking Defenses}\label{ssec:breaking-confidence-masking}
Answering question 2), we showcase an example where our label-only attacks \emph{outperform} prior attacks by a significant margin, despite the strictly more restricted query interface that they assume.
We evaluate 
\emph{defenses} against \mi attacks and show that while these defenses do protect against existing confidence-vector attacks, they have little to no effect on our label-only attacks.
Because any ML classifier providing confidences also provides the predicted labels, our attacks fall within their threat model, refuting these defenses' security claims.

We identify a common pattern to these defenses that
we call \emph{confidence masking}, wherein defenses aim to prevent \mi by directly minimizing the privacy leakage in a model's confidence scores. 
To this end, 
confidence-masking defenses explicitly or implicitly mask (or, obfuscate) the information contained in the model's confidences, (e.g., by adding noise) to thwart existing attacks. 
These defenses, however, \emph{have a minimal effect on the model's predicted labels}. 
MemGuard~\cite{memguard} and prediction purification~\cite{yang2020predictionpurification} explicitly maintain the invariant that the model's predicted labels are not affected by the defense, i.e., 
\[ \forall\ x, \quad \argmax h(x) = \argmax h^{\text{defense}}(x) \;, \]
where $h^{\text{defense}}$ is the defended version of the model $h$.

An immediate issue with the design of confidence-masking defenses is that,
by construction, they will not prevent any label-only attack.
Yet, these defenses were reported to drive the success rates of existing \mi attacks to within chance. This result suggests that prior attacks fail to properly extract membership
information contained in the
model's 
predicted labels, and 
implicitly contained within its scores. 
Our label-only attack performances clearly indicate that confidence masking is not a viable defense strategy against \mi.

We show that two peer-reviewed defenses, MemGuard~\cite{memguard} and adversarial regularization~\cite{adversarialregularization},
fail to prevent label-only attacks, and thus, do not significantly reduce \mi compared to an undefended model.
Other 
proposed defenses, e.g., reducing the precision or cardinality of the confidence-vector~\cite{cornellMI,demystify,salem2018mlleaks}, and recent defenses like prediction purification~\cite{yang2020predictionpurification}, also rely on confidence masking: they are unlikely to resist label-only attacks. See Supplement \S~\ref{app:confidence-masking} for more details on these defenses.

\subsection{Breaking MemGuard}\label{ssec:breaking-memguard}
We implement the strongest version of MemGuard that can make \emph{arbitrary} changes to the confidence-vector while leaving the model's predicted label unchanged. 
Observing Figure~\ref{fig:labelonly} and Table~\ref{tab:cifar100} (b) and (d), we see that MemGuard successfully defends against prior confidence-vector attacks, but as expected, offers no protection against our label-only attacks. All our attacks significantly outperform the (non-adaptive) confidence-vector and the baseline gap attack.

The main reason that~\citet{memguard} found MemGuard to protect against confidence-vector attacks is because these attacks were not properly \emph{adapted} to this defense. Specifically, MemGuard is evaluated against confidence-vector attacks that are tuned on source models \emph{without MemGuard enabled}. This observation also holds for other defenses such as~\citet{yang2020predictionpurification}. Thus, these attacks' membership predictors are tuned to distinguish members
from non-members based on high confidence scores, which these defenses obfuscate. In a sense, a label-only attack like ours is the ``right'' adaptive attack against these defenses: since the model's confidence scores are no longer reliable, the adversary's best strategy is to 
use
hard labels, which these defenses explicitly do not modify. Moving forward, we recommend that the trivial gap baseline serve as an indicator of confidence masking: a confidence-vector attack should not perform significantly worse than the gap attack for a defense to protect against \mi. 
Thus, to protect against (all) \mi attacks, a defense cannot solely post-process the confidence-vector---the model will still be vulnerable to label-only attacks.

\begin{figure*}[t]
    \centering
    \subfloat[Rotation attack]{\includegraphics[width=.7\columnwidth]{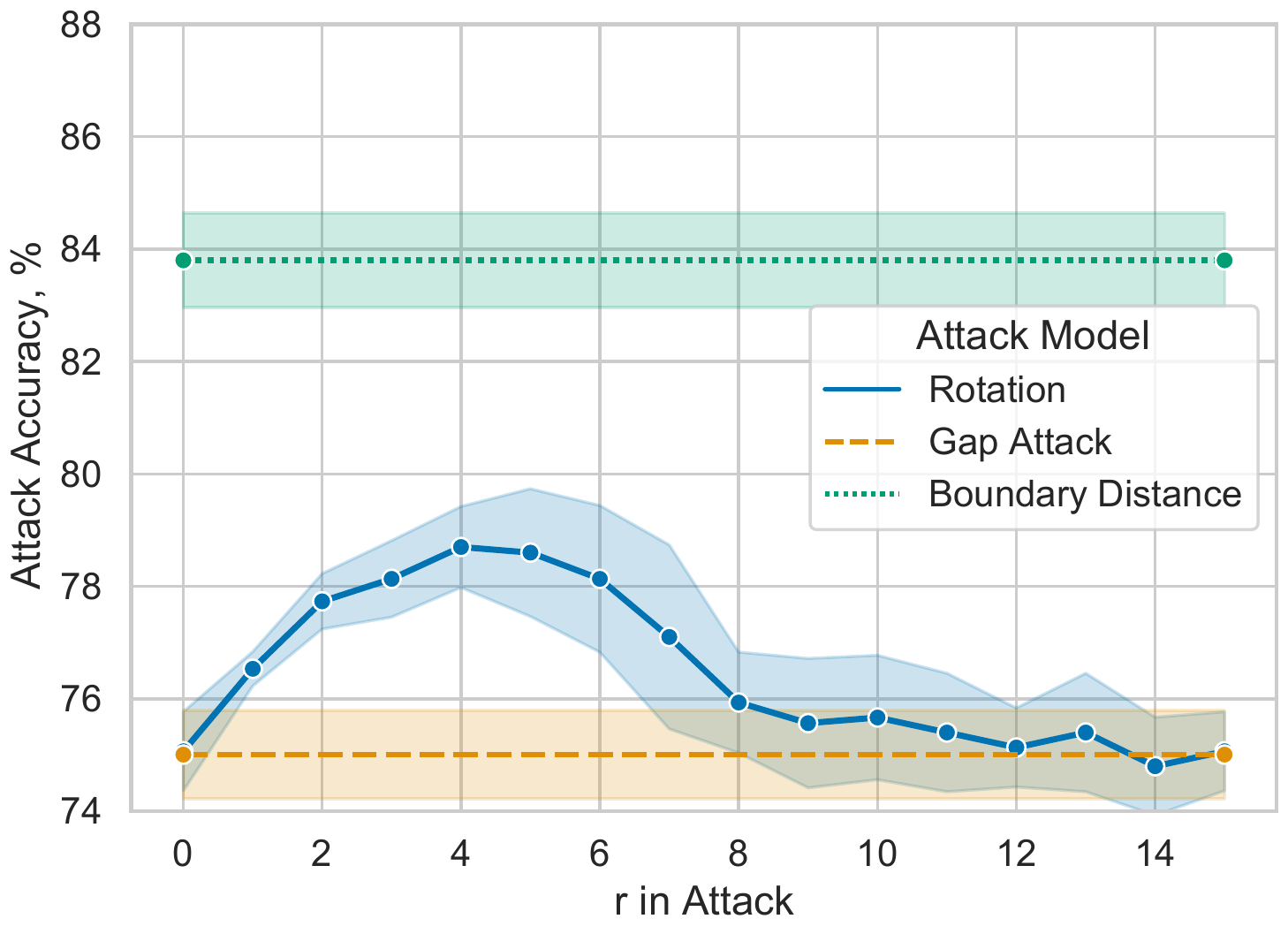}}
    \subfloat[Translation attack]{\includegraphics[width=.7\columnwidth]{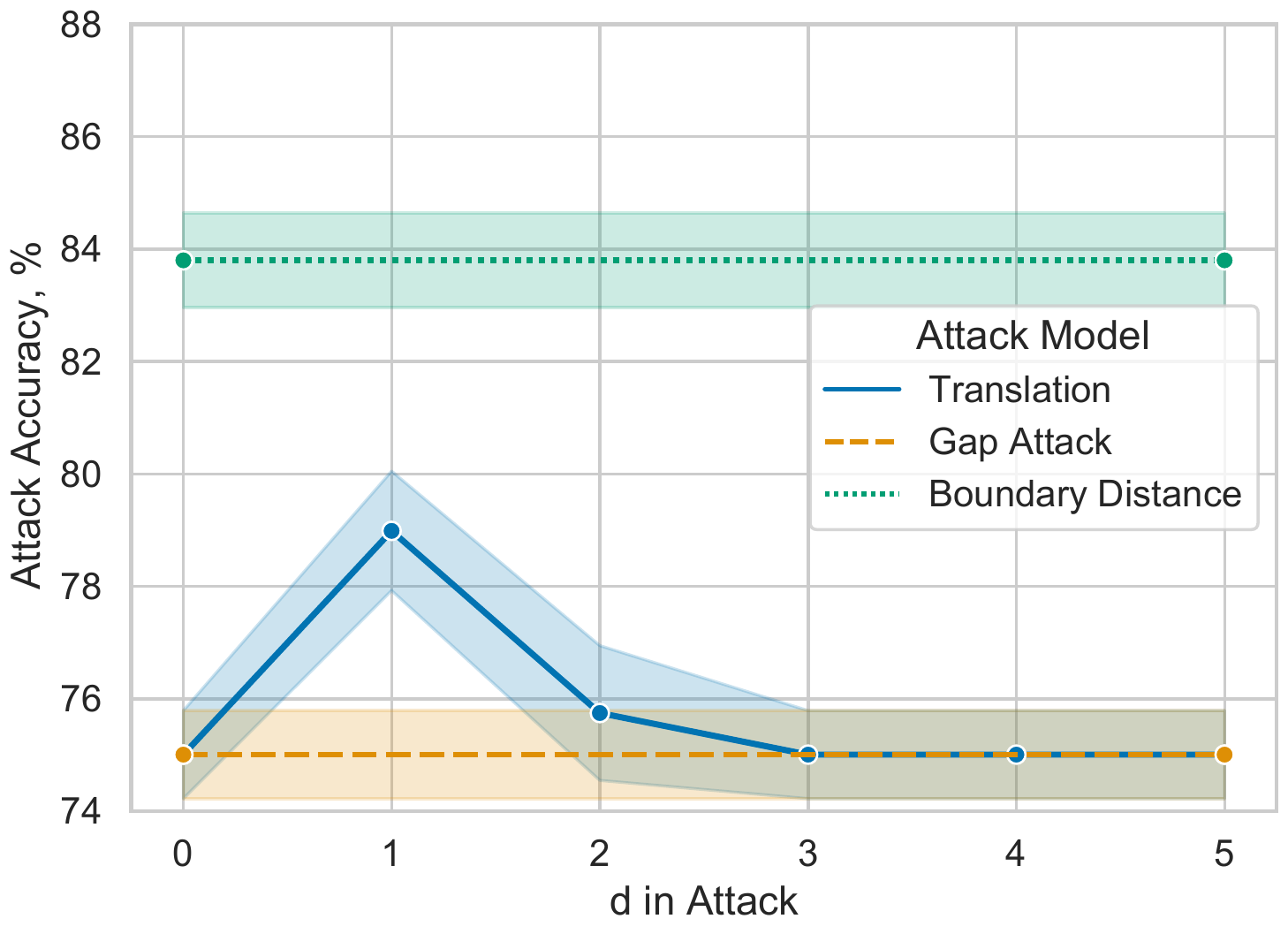}}
    \subfloat[Boundary distance attack]{\includegraphics[width=.65\columnwidth]{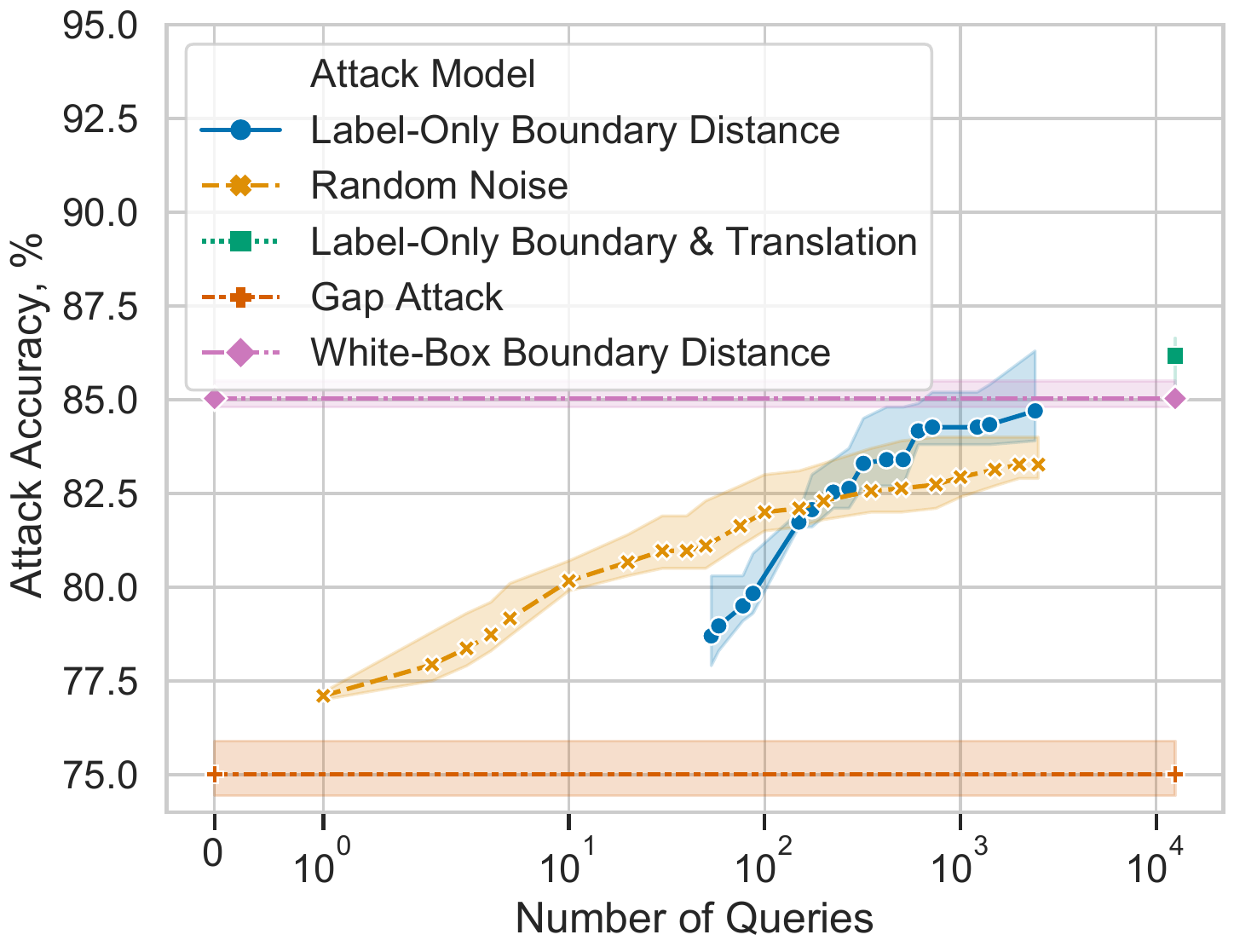}}
    \vspace{-0.25em}
    \caption{
    \textbf{Comparison of query settings for different label-only \mi attacks on CIFAR-10}. Target model are trained on a subset of $2{,}500$ data points. In (a) and (b), we compare the performance of the data augmentation attack against two baselines (the gap attack and the label-only boundary distance attack, with $2,500$ queries), as we increase the $r$ and $d$ parameters.
    In (c), we compare attacks that threshold on a point's distance to the boundary in a black-box setting with a white-box baseline using Carlini and Wagner's attack~\cite{carliniwagner}.
    We describe these attacks in \S~\ref{ssec:decisionboundarydistance}.
    Costs are $\approx\$0.1$ per $1000$ queries\textsuperscript{\ref{fnote:pricing}}. 
    }
    \label{fig:augattackvary}
    \vspace{-1em}
\end{figure*}

\subsection{Breaking Adversarial Regularization}\label{ssec:breakingadversarialregularization}
The work of~\citet{adversarialregularization} differs
from MemGuard and prediction purification in that it does not simply obfuscate confidence vectors at test time. Rather, it jointly trains a target model and a defensive confidence-vector \mi classifier in a min-max fashion: the attack model to maximize \mi and the target model to produce accurate outputs that yet fool the attacker. See Supplement \S~\ref{app:adversarial-regularization} for more details.

We train a target model defended using adversarial regularization, exactly as in~\cite{adversarialregularization}.
By varying its hyper-parameters, we achieve a defended state where the confidence-vector attack is within $3$ percentage points of chance, as shown in Supplement Figure~\ref{fig:adversarialregularization}. Again, our label-only attacks significantly outperform this attack (compare Figures~\ref{fig:defensesattackacc} (a) and (b)) because the train-test gap is only marginally reduced; 
this defense is not entirely ineffective---it prevents label-only attacks from exploiting beyond $3$ percentage points of the gap attack.
However, when label-only attacks are sufficiently defended, it achieves significantly worse test accuracy trade-offs than other defenses (see Figure~\ref{fig:defensestestacc}).
And yet, evaluating the defense solely on
confidence-vector attacks overestimates the achieved privacy.

\subsection{The Query Complexity of Label-Only Attacks}
\label{ssec:label-only-query}

We now answer question 3) and investigate how the query budget affects the success rate of different label-only attacks.

Recall that our rotation attack evaluates $N=3$ queries of images rotated by $r^\circ$ and our translation attack $N=4d+1$ for shifts satisfying a total displacement of $d$.
Figure~\ref{fig:augattackvary} (a)-(b) shows
that there is a 
range of perturbation magnitudes for which the attack exceeds the baseline (i.e., $1 \leq r \leq 8$ for rotations, and $ 1 \leq d \leq 2$ for translations). When the augmentations are too small or too large, the attack performs poorly because the augmentations 
have a similar effect on both train and test samples 
(\ie small augmentations rarely change model predictions and large augmentations often cause misclassifications).
An optimal parameter choice ($r$ and $d$) outperforms the baseline by $3$-$4$ percentage-points, which an adversary can tune using its local source model. 
As we will see in 
\S~\ref{sec:generalization}, these attacks outperform all others
on models that used data augmentation \emph{at training time}.

\begingroup
\setmuskip{\thickmuskip}{0mu}
In Figure~\ref{fig:augattackvary} (c), we compare different boundary distance attacks, discussed in \S~\ref{ssec:decisionboundarydistance}.
With $\approx2{,}500$ queries, the label-only attack matches the white-box upper-bound
using $\approx2{,}000$ queries and also matches the best confidence-vector attack (see Figure~\ref{fig:labelonly}). With $\approx12{,}500$ queries, our combined attack can outperform all others. 
Query limiting would likely not be a suitable defense, as Sybil attacks~\cite{douceur2002sybil} can circumvent it; even in low query regimes ($<100$) our attacks outperform the trivial gap by $4$ percentage points. Finally, with $<300$ queries, our simple noise robustness attack outperforms our other label-only attacks. At large query budgets, our boundary distance attack produces more precise distance estimates and outperforms it. Note that the monetary costs are modest at $\approx\$0.25$ per sample\footnote{\label{fnote:pricing}\url{https://www.clarifai.com/pricing}}.
\endgroup

%% file: content/5_new_defenses.tex
\section{Defending with Better Generalization}
\label{sec:generalization}
Since confidence-masking defenses cannot robustly defend against \mi attacks, we now investigate to what extend standard regularization techniques---that aim to limit a models' ability to overfit to its training set---can. 
We study how data augmentation, transfer learning, dropout~\cite{srivastava2014dropout}, $\ell_1$/$\ell_2$ regularization, and differentially private (DP) training~\cite{dpdeeplearning} impact \mi.

We explore three questions in this section:
\renewcommand{\theenumi}{\Alph{enumi}}
\begin{enumerate}[topsep=0pt,itemsep=-1ex,partopsep=1ex,parsep=1ex]
    \item How does training with data augmentation impact \mi attacks, especially those that evaluate augmented data?
    \item How well do other standard machine learning regularization techniques 
    help in reducing \mi?
    \item How do these defenses compare to differential privacy, which can provide formal guarantees against any form of membership leakage?
\end{enumerate}

\subsection{Training with Data Augmentation Exacerbates \mi}
\label{ssec:augmentations}

Data augmentation is commonly used in machine learning to reduce overfitting and encourage generalization, especially in low data regimes~\cite{dataaugmentsurvey, healthaugment}.
 Data augmentation is an interesting case study for our attacks. As it reduces a model's overfitting, one would expect it to reduce \mi. 
But, a model trained with augmentation will have been trained to strongly recognize 
$x$
and its augmentations, which is precisely the signal that our data augmentation attacks exploit.

We train target models with data augmentation similar to \S~\ref{ssec:augmentation_attack_design} and focus on translations as they are most common in computer vision. We use a simple pipeline where all translations of each image is evaluated in a training epoch. Though this differs slightly from the standard random
sampling, we choose it to illustrate the maximum \mi when the adversary's queries exactly match the samples seen in training. 

\begin{figure}[t]
    \centering
    \includegraphics[width=0.85\columnwidth]{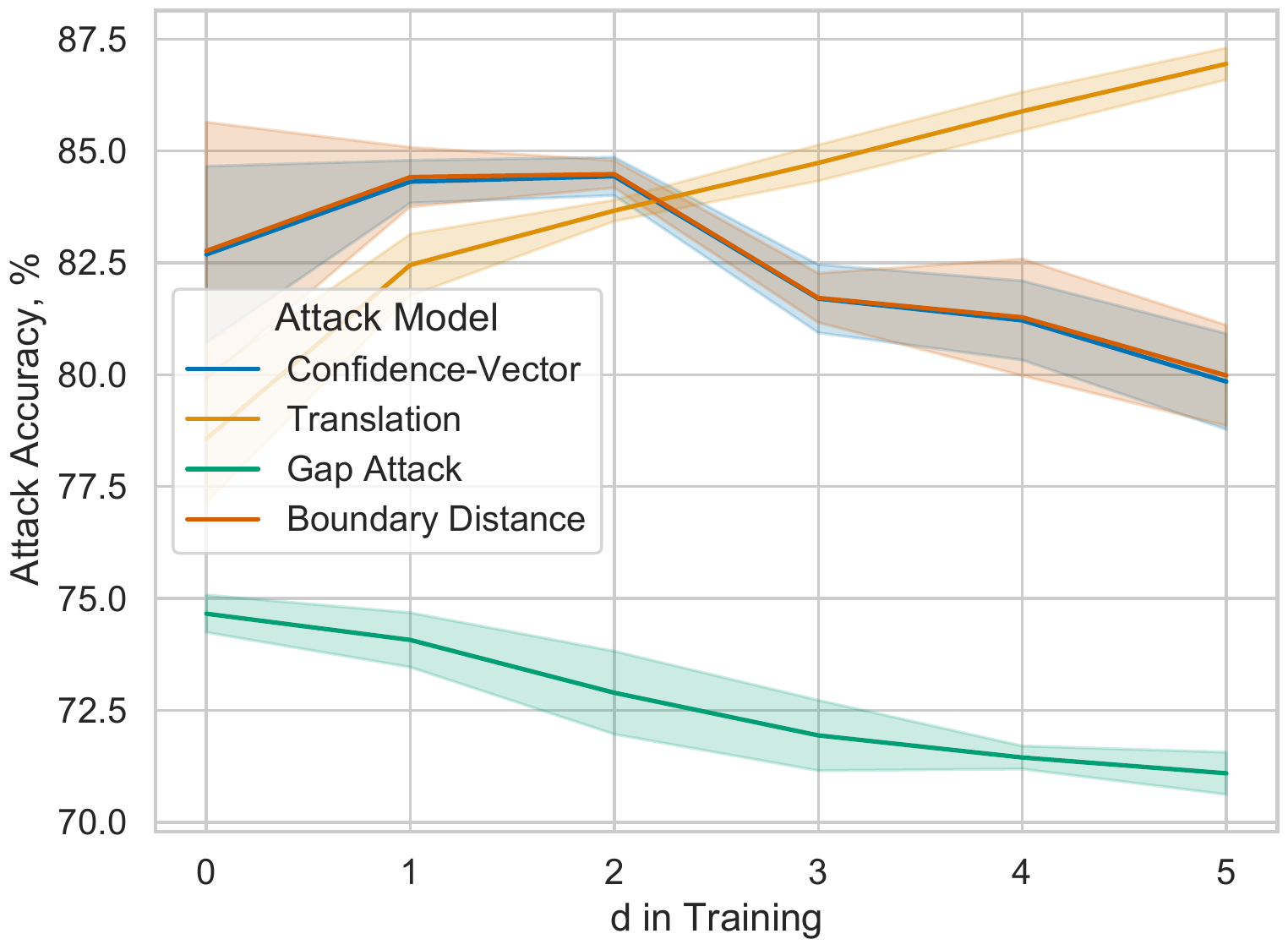}
    \vspace{-0.5em}
    \caption{\textbf{Accuracy of \mi attacks on CIFAR-10 models trained with data augmentation}
    on a subset of 2500 images.
    As in our attack, $d$ controls the number of pixels by which images are translated during training, where no augmentation is $d=0$.
    For models trained with significant amounts of data augmentation, \mi attacks become \emph{stronger} despite 
    it generalizing better. 
    }
    \vspace{-6pt}
    \label{fig:rotateaugmentdefense}
\end{figure}

Observe from Figure~\ref{fig:rotateaugmentdefense} that augmentation reduces overfitting and improves generalization: test accuracy increases from $49.7\%$ without translations to $58.7\%$ at $d=5$ and the train-test gap decreases. Due to improved generalization, the confidence-vector and boundary distance attacks' accuracies \emph{decrease}. 
\emph{Yet, the success rate of the data augmentation attack increases.}
This increase
confirms our initial intuition that the model now leaks \emph{additional} membership information via its invariance to training-time augmentation. 
Though 
the model trained with $d=5$ pixel shifts has higher 
test accuracy, our data augmentation attack exceeds the confidence-vector performance on the 
non-augmented model.\footnote{Though we find in Supplement Figure~\ref{fig:choosed} that the attack is strongest when the adversary correctly guesses $d$, we note that these values are often fixed for a domain and image resolution. Thus, adversarial knowledge of the augmentation pipeline is not a strong assumption.}
Thus, 
model generalization is not the only variable affecting its membership leakage: models that overfit less on the original 
data
may actually be more vulnerable to \mi \emph{because they implicitly overfit more on a related 
dataset}.

\paragraph{Attacking a high-accuracy ResNet}
We use, without modification, the pipeline from FixMatch~\cite{fixmatch},
which trains a ResNet-28 to $96\%$ accuracy
on the CIFAR-10 dataset, comparable to the state of the art. As with our other experiments, 
this model is trained using a subset of CIFAR-10, which sometimes leads to observably overfit models indicated by a higher gap attack accuracy. We train models using four regularizations, all random: vertical flips, shifts by up to $d=4$ pixels, image cutout~\cite{cutout}, and (non-random) weight decay of magnitude $0.0005$. All are either enabled 
or disabled. 

\begin{figure}[h!]
    \centering
    \subfloat[Without Augmentations]{\includegraphics[width=.9\columnwidth]{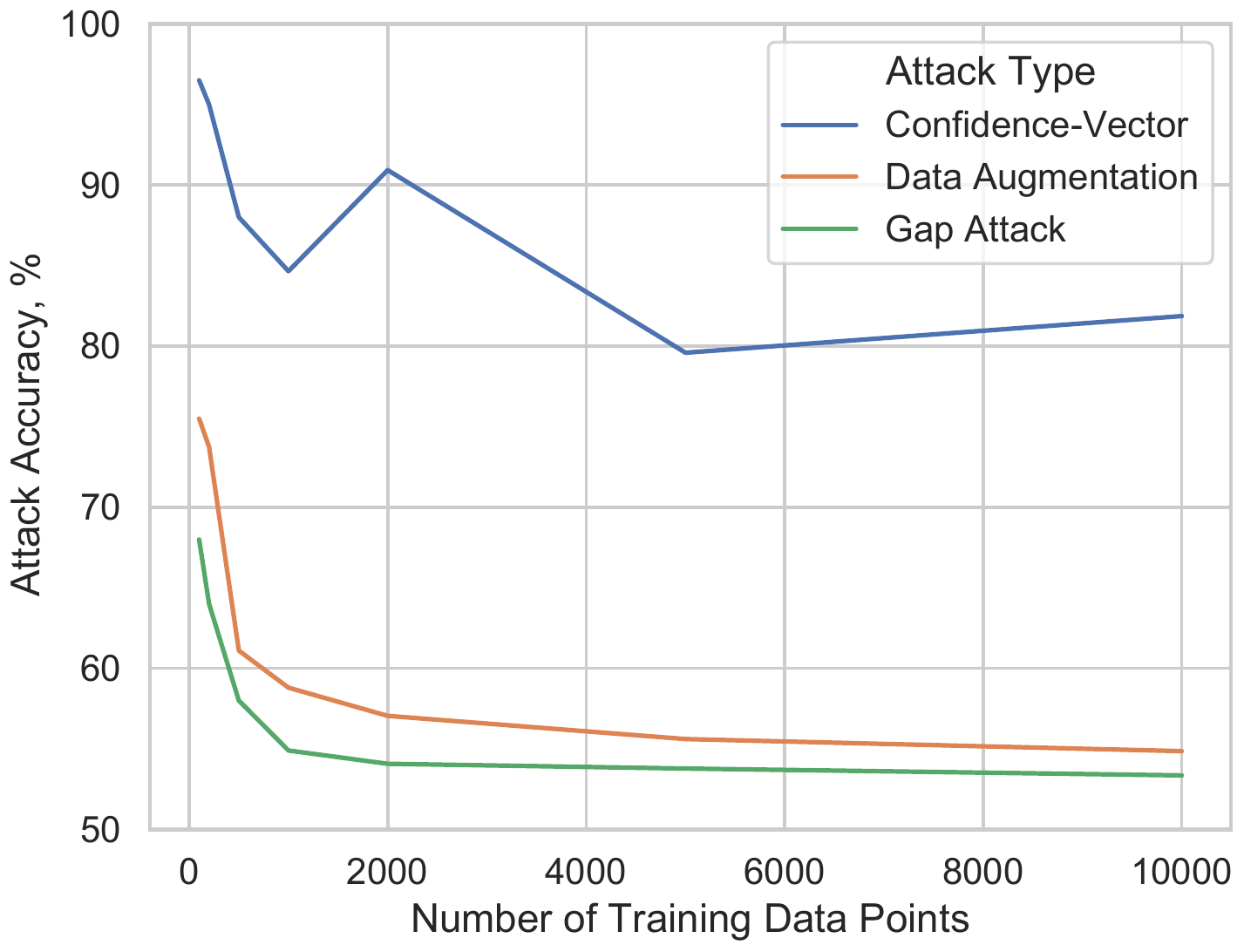}}\vspace{-1.em}\\
    \subfloat[With Augmentations]{\includegraphics[width=.9\columnwidth]{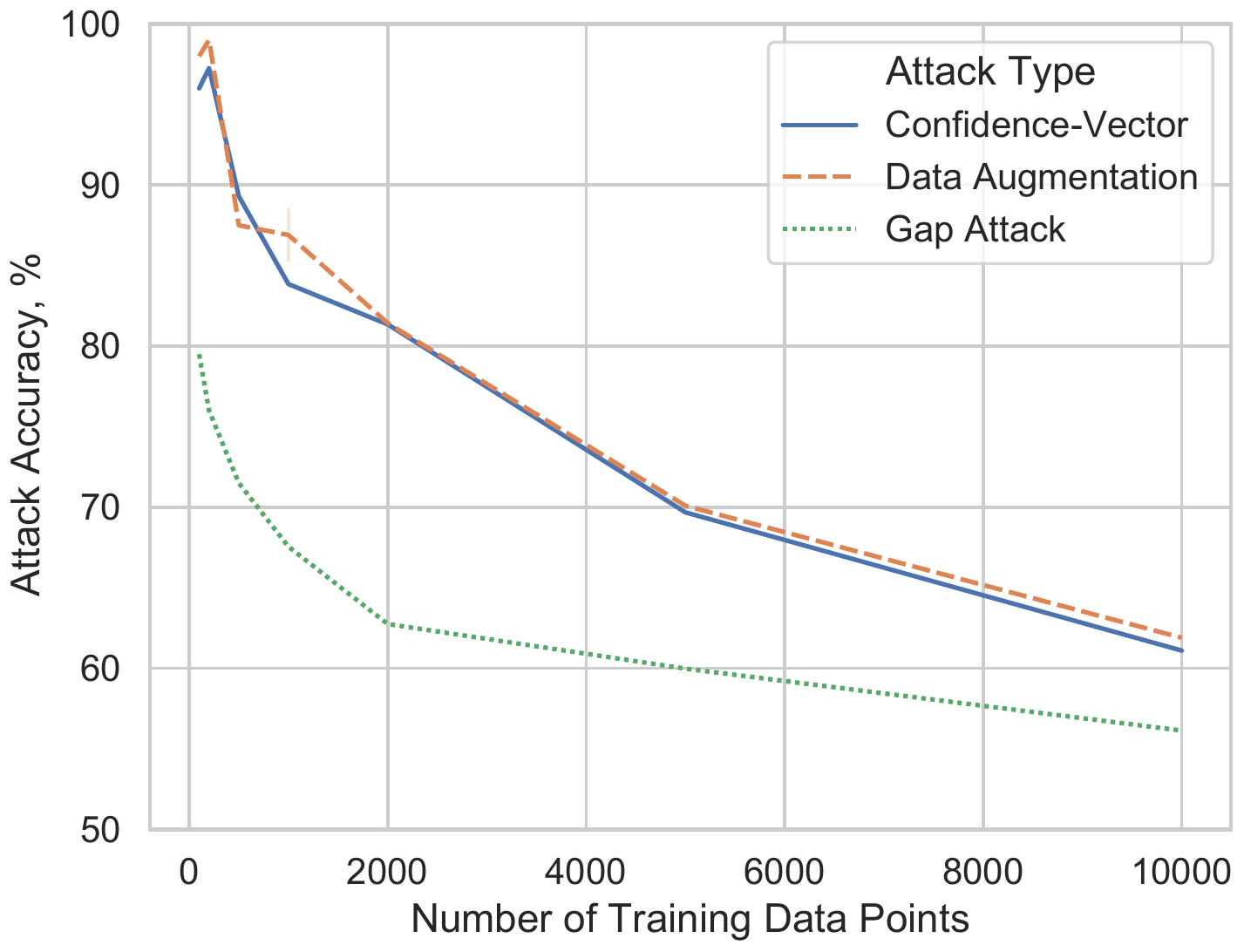}}
    \vspace{-0.5em}
    \caption{\textbf{Accuracy of membership inference attacks on CIFAR-10 models trained as in FixMatch~\cite{fixmatch}.} Our data augmentation attacks, which mimic the training augmentations, match or outperform the confidence-vector attacks when augmentations were used in training. We evaluate $1000$ randomly generated augmentations for this attack. As in previous experiments, we find our label-only distance attack performs on par with the confidence-vector attack. Interestingly, the gap attack accuracy also improves due to a relatively larger increase in training accuracy. ``With Augmentations'' and ``Without Augmentations'' refer to using all regularizations, as in FixMatch, or none, respectively.
    }
    \label{fig:fixmatchdefense}
\end{figure}

Our results here, shown in Figure~\ref{fig:fixmatchdefense} corroborate those obtained with the simpler pipeline above: \emph{though test accuracy improves, our data augmentation attacks match or outperform the confidence-vector attack}.

\renewcommand{\theenumi}{\arabic{enumi}}

\subsection{Other Techniques to Prevent Overfitting}\label{ssec:other-techniques-to-preventing-overfitting}
We explore questions B)-C) 
using
other standard regularization techniques,
with details in Supplement~\ref{apx:regularizers}. In transfer learning, we either 
only
train a new last layer
(\emph{last layer fine-tuning}), or fine tune the entire model (\emph{full fine-tuning}).

\begin{figure}[h!]
    \centering
    \includegraphics[width=.9\columnwidth]{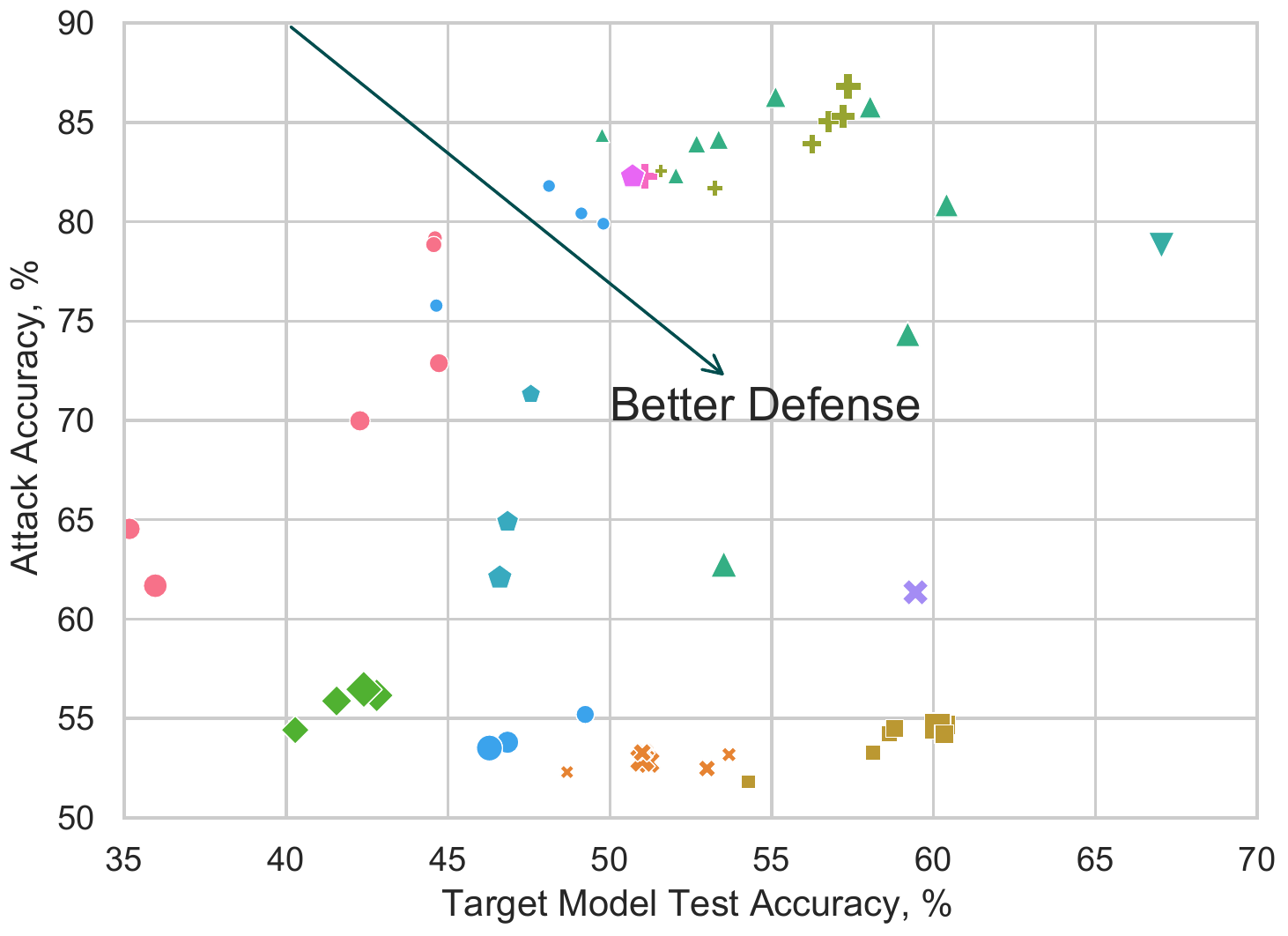}
    \vspace{-1em}
    \caption{\textbf{Test accuracy and label-only attack accuracy for 
    various defenses.
    } 
    Same setup as Figure~\ref{fig:defensesattackacc}. Models towards the bottom right are more private and 
    more 
    accurate.
    \vspace{-1em}}
    \label{fig:defensestestacc}
\end{figure}

We pre-train a model on CIFAR-100 to $51.6\%$ test accuracy and then use transfer learning. We find that boundary distance attack performed on par with the confidence-vector in all cases. We observe that last layer fine-tuning degrades all our attacks to the generalization gap, preventing non-trivial \mi (see Figure~\ref{fig:transferlearn} in Supplement \S~\ref{app:additional-figures}). This result corroborates intuition: linear layers have less capacity to overfit compared to neural networks. We observe that full fine-tuning leaks more membership inference but achieves better test accuracies, as shown in Figure~\ref{fig:defensestestacc}.

Finally, DP training~\cite{dpdeeplearning}
formally enforces that the trained model does not strongly depend on any individual training point---that it does not overfit. 
We use differentially private gradient descent (DP-SGD)~\cite{dpdeeplearning} (see Supplement \S~\ref{apx:regularizers}). 
To achieve comparable test accuracies as undefended models, the formal privacy guarantees become mostly meaningless (i.e., $\epsilon>100$).

\begin{figure}[h!]
    \centering
    \captionsetup[subfigure]{}
    \subfloat[Label-Only Attacks]{\includegraphics[width=.85\columnwidth]{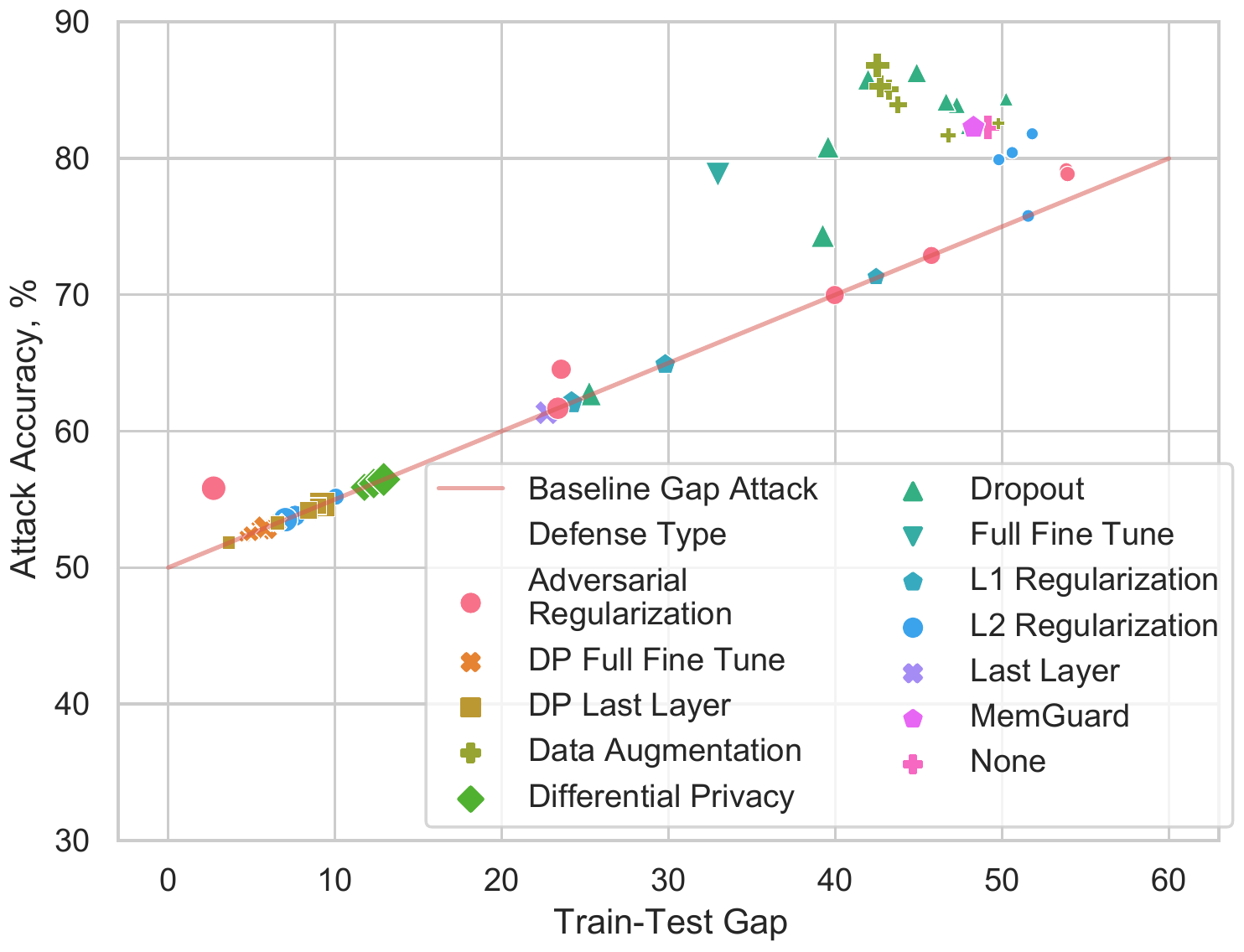}}\\[-0.2em]
    \subfloat[Full Confidence-Vector Attacks]{\includegraphics[width=.85\columnwidth]{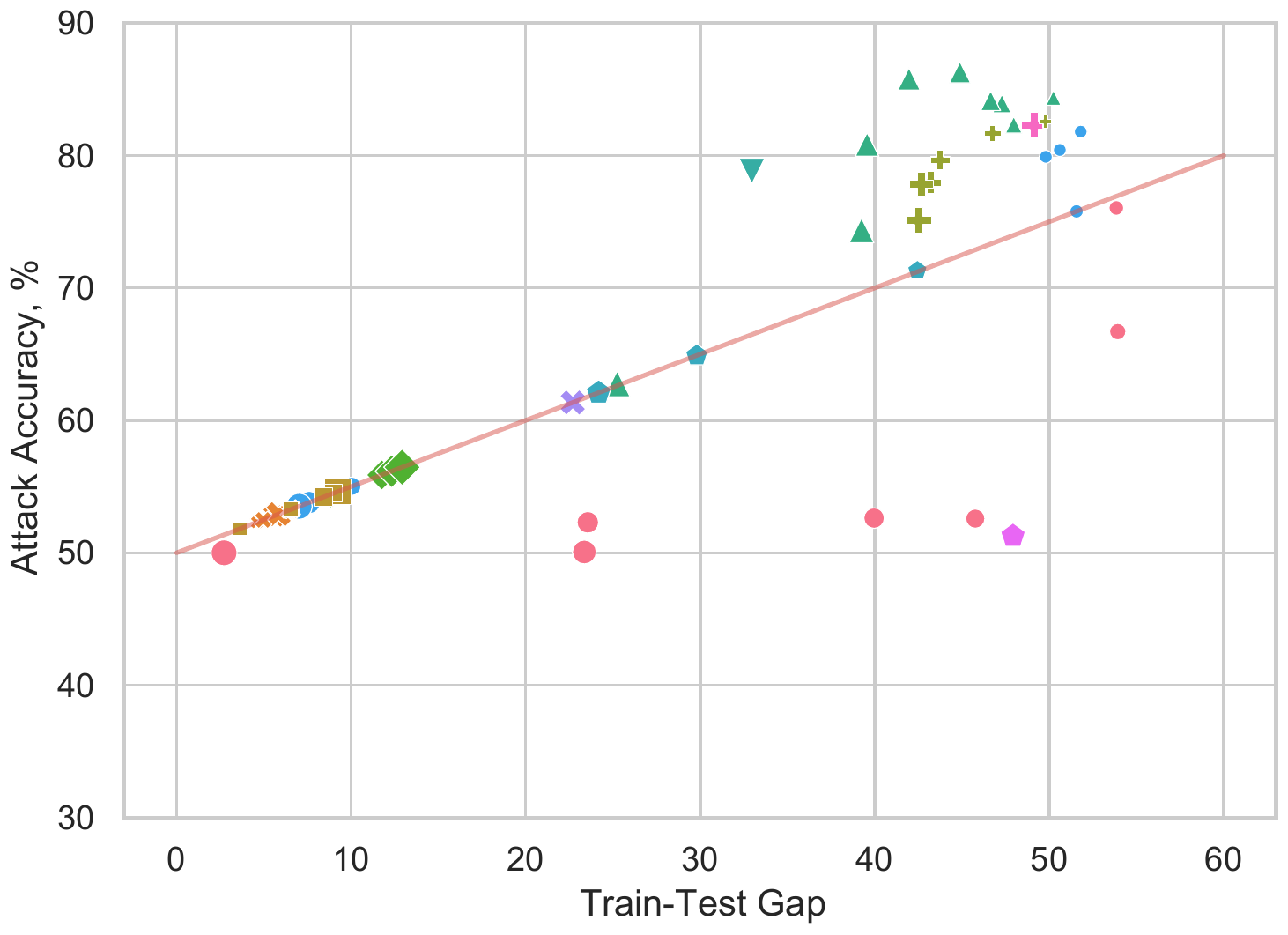}}
    \caption{\textbf{Comparison of \mi attacks on various defenses.} Target models are trained on 2500 data points from CIFAR-10. Point sizes indicate relative regularization amounts within a defense.
    \vspace{-1em}
    }
    \label{fig:defensesattackacc}
\end{figure}

In Figure~\ref{fig:defensesattackacc}, we find that most forms of regularization fail to prevent even the baseline gap attack from reaching $60\%$ accuracy or more. Only strong $\ell_2$ regularization ($\lambda \geq 1$) and DP training consistently reduced \mi. Figure~\ref{fig:defensestestacc} gives us a best understanding of the privacy-utility trade-off. We see that both prevent \mi at a high cost in test-accuracy---they cause the model to \emph{underfit}. However, we also clearly see the utility benefits of transfer learning: these models achieve consistently better test-accuracy due to features learned from non-private data. Combining DP training with transfer learning mitigates privacy leakage at only minimal cost in test accuracy, achieving the best tradeoff. When transfer learning is not an option, dropout performs better.

%% file: content/6_outlier_analysis.tex
\section{Worst-Case (Outlier) \mi}\label{sec:outlier}
Here, we perform \mi only for a small subset of ``outliers''. Even if a model generalizes well on average, it might still have overfit to unusual
data
in the tails of the distribution~\cite{secretsharer}. 
We use a similar but modified process as~\citet{generalizedmi} to identify potential outliers.

First, the adversary uses a source model $\hat{h}$ to map each targeted data point, $x$, to its feature space, or the activations of its penultimate layer, denoted as $z(x)$. 
We define two points $x_1, x_2$ as \emph{neighbors} if their features are close, i.e., $d(z(x_1), z(x_2)) \leq \delta$, where $d(\cdot, \cdot)$ is the cosine distance and $\delta$ is a tunable parameter.
An outlier is a point with less than $\gamma$ neighbors in 
$z(x)$
where $\gamma$ is another tunable parameter. Given a dataset $X$ of potential targets and an intended fraction of outliers $\beta$ (e.g., $1\%$ of $X$), we tune $\delta$ and $\gamma$ so that a $\beta$-fraction of points $x \in X$ are 
outliers. We use precision as the \mi success metric.

We run our attacks on the outliers of the same models as in Figure~\ref{fig:defensesattackacc}. We find in See Figure~\ref{fig:outliers} in Supplement Section~\ref{app:additional-figures}, that we can always improve the attack by targeting outliers, but that strong $L2$ regularization and DP training prevent \mi. As before, we find that the label-only boundary distance attack matches the confidence-vector attack performance.

%% file: content/7_discussion_conclusion.tex
\section{Conclusion}

We developed three new label-only membership inference attacks 
that can match, and even exceed, the success of prior confidence-vector attacks, despite operating in a more restrictive adversarial model.
Their label-only nature requires fundamentally different attack strategies, that---in turn---cannot be trivially prevented by obfuscating a model's confidence scores.
We have used these attacks to break two state-of-the-art defenses to membership inference attacks. 

We have found that the problem with these ``confidence-masking'' defenses runs deeper: they cannot prevent any label-only attack.
As a result, any defenses against \mi necessarily have to help reduce a model's train-test gap.

Finally, via a rigorous evaluation across many proposed defenses to \mi, we have shown that differential privacy (with transfer learning) 
provides the strongest defense,
both in an average-case and worst-case sense, but that it may come at a cost in the model's test accuracy.

To center our analysis on comparing the confidence-vector and label-only settings, we use the same threat model as prior work~\cite{cornellMI} and leave a fine-grained analysis of label-only attacks under reduced adversarial knowledge
(e.g., reduced data and model architecture knowledge~\cite{privacyoverfittingmi,salem2018mlleaks}) to future work.

%% file: content/8_appendix.tex
\section{Background}
\label{sec:problemdefinition}

\subsection{Machine Learning}
We consider supervised classification tasks~\cite{mlp, mltheoryalgorithms}, wherein a model is trained to predict some class label $y$, given input data $x$.
Commonly, $x$ may be an image or sentence and $y$ is then the corresponding label, e.g., a digit 0-9 or a text sentiment.

We focus our study on neural networks~\cite{deeplearning}: functions composed as a series of linear-transformation layers, each followed by a non-linear activation. 
The overall layer structure is called the model's \emph{architecture} and the learnable parameters of the linear transformations are the \emph{weights}. 
For a classification problem with $K$-classes, the last layer of a neural network outputs a vector $\textbf{v}$ of $K$ values (often called logits). The \emph{softmax} function is typically used to convert the logits into normalized confidence scores:\footnote{While it is common to refer to the output of a softmax as a ``probability vector'' because its components are in the range $[0,1]$ and sum to $1$, we refrain from using this terminology given that the scores output by a softmax cannot be rigorously interpreted as probabilities~\cite{gal2016uncertainty}}. $\text{softmax}(\textbf{v})_i \coloneqq {e^{v_i}}/{\sum_{i=1}^K e^{v_i}} \in [0, 1]$.
For a model $h$,  we define the model's output $h(x)$ as the vector of softmax values. The model's predicted label is the class with highest confidence, \ie $\text{argmax}_i\ h(x)_i$.

\subsubsection{Data Augmentation}
Augmentations are natural transformations of existing data points that preserve class semantics (e.g., small translations of an image), which are used 
to improve the generalization of a classifier~\cite{autoaugment,fixmatch,cvaugment}. They are commonly used on  state-of-the-art models~\cite{heresnet,autoaugment,perezeffectivenessaugment}
to increase the diversity of the finite training set,
 without 
the need to acquire 
more labeled data (in a costly process). Augmentations are especially important in low-data regimes~\cite{sajjad2019multi,augmentnlplowresource,augmentacousticlowresource} and are domain-specific: they apply to a certain type of input, (e.g., images or text).

We focus on image classifiers, where the main types of augmentations are affine transformations (rotations, reflections, scaling, and shifts), contrast adjustments, cutout~\cite{cutout}, and blurring (adding noise). By synthesizing a new data sample as an augmentation of an existing data sample, $x' = \augment(x)$, the model can learn a more semantically-meaningful set of features. Data augmentation can potentially teach the machine learning model to become invariant to the augmentation (e.g., rotationally or translationally invariant).

\subsubsection{Transfer Learning}
Transfer learning is a common technique used to improve generalization in low-data regimes~\cite{tan2018surveytransferlearn}. By leveraging data from a \emph{source task}, it is possible to transfer knowledge to a \emph{target task}. Commonly, a model is trained on the data of the source task and 
then fine-tuned on data from the output task.
In the case of neural networks, it is common to fine-tune either the entire model or just the last layer.

\subsection{Membership Inference}

Membership inference attacks~\cite{cornellMI} are a form of privacy leakage that identify if a given data sample was in a machine learning model's training dataset. 
Given a sample $x$ and access to a trained model $h$, the adversary uses a classifier or decision rule $f_h$ to compute a membership \emph{prediction} $f(x; h) \in \{0, 1\}$, with the goal that $f(x; h) = 1$ whenever $x$ is a training point. The main challenge in mounting a membership inference attack is creating the classifier $f$, under various assumptions about the adversary's knowledge of $h$ and its training data distribution.

Prior work assumes that an adversary has only black-box access to the trained model $h$, via a query interface that on input $x$ returns part or all of the confidence vector $h(x)$.

\paragraph{Shadow Models}\label{ssec:shadowmodels}
The original membership inference attack of Shokri \etal~\cite{cornellMI} creates a membership classifier $f(x; h)$, tuned on a number of local ``shadow'' (or, source) models. Assuming the adversary has access to data from the same (or similar) distribution as $h$'s training data, the shadow model approach trains the auxiliary source models $\hat{h}_i$ on this data. Since $\hat{h}_i$ is trained by the adversary, they know whether or not any data point was in the training set, and can thus construct a dataset of confidence vectors $\hat{h}_i$ with an associated membership label $m \in \{0,1\}$. The adversary trains a classifier $f$ to predict $m$ given $\hat{h}_i(x)$. Finally, the adversary queries the targeted model $h$ to obtain $h(x)$ and uses $f$ to predict the membership of $x$ in $h$'s training data.

Salem \etal~\cite{salem2018mlleaks} later showed that this attack strategy can succeed even without data from the same distribution as $h$, and only with data from a similar task (e.g., a different vision task). They also showed that training shadow models is unnecessary: applying a simple threshold predicting 
$f(x; h) = 1$ ($x$ is a member)
when
the max prediction confidence,
$\max_i h(x)$,
is above a tuned threshold, suffices.

\paragraph{Towards Label-only Approaches} 
 Yeom \etal~\cite{privacyoverfittingmi} propose a simple baseline attack: the adversary predicts a data point $x$ as being a member of the training set when $h$ classifies $x$ correctly. The accuracy of this baseline attack directly reflects the gap in the model's train and test accuracy: if $h$ overfits (i.e., obtains higher accuracy) on its training data 
, this baseline attack will achieve non-trivial membership inference. We call this the gap attack. 
If the adversary's target points are equally likely to be members or non-members of the training set (see Appendix~\ref{ssec:on-measuring-success})
, this attack achieves an accuracy of
\begin{equation*}
    1/2 + (\text{acc}_{\text{train}} - \text{acc}_{\text{test}})/2 \;,
\end{equation*}
where $\text{acc}_{\text{train}}, \text{acc}_{\text{test}} \in [0, 1]$ are the target model's accuracy on training data and held out data respectively.

To the best of our knowledge, this is the only attack proposed 
in prior work that makes use of only the model's predicted label, $y = \argmax_i h(x)_i$. Our goal is to investigate how this simple baseline can be surpassed to achieve label-only membership inference attacks that perform on par with attacks that use access to the model's confidence scores. 
\paragraph{Indirect Membership Inference}
The work of Long \etal~\cite{generalizedmi} investigates 
membership inference through \emph{indirect access}, wherein 
the adversary only queries $h$ on inputs $x'$ that are related to $x$, but not $x$ directly.
Our label-only attacks similarly make use of information 
gleaned from querying $h$ on data points related to $x$ (specifically, perturbed versions of $x$).

The main difference is that we focus on label-only attacks, whereas the work of Long \etal~\cite{generalizedmi} assumes adversarial access to the model's confidence scores. Our attacks will also be allowed to query and obtain the label at the chosen point $x$.
\paragraph{Adversarial Examples and Membership Inference}
Song \etal~\cite{song2019privacy} also make use of adversarial examples to infer membership. Their approach crucially differs from ours in two aspects: (1) they assume access to and predict membership using the confidence scores, and (2) they target models that were explicitly trained to be robust to adversarial examples. 
In this sense, (2) bares some similarities with our attacks on models trained with data augmentation (see Section
~\ref{sec:generalization}, where we also find that a model's invariance to some perturbations can leak additional membership signal).
\paragraph{Defenses} 

Defenses against membership inference broadly fall into two categories. 

First, standard regularization techniques, such as L2 weight normalization~\cite{cornellMI,memguard,demystify,adversarialregularization}, dropout~\cite{memguard}, or differential privacy have been proposed to address the role that overfitting plays in a membership inference attack's success rate~\cite{cornellMI}. Heavy regularization has been shown to limit overfitting and to effectively defend against membership inference, but may result in a significant degradation in the model's accuracy. Moreover, Yeom \etal~\cite{privacyoverfittingmi} show that overfitting is sufficient, but not necessary, for membership inference to be possible.

Second, defenses may reduce the information contained in a model's confidences, e.g., by truncating them to a lower precision~\cite{cornellMI}, reducing the dimensionality of the confidence-vector to only some top $k$ scores~\cite{cornellMI,demystify}, or perturbing confidences via an adversary-aware ``minimax'' approach~\cite{adversarialregularization,yang2020predictionpurification,memguard}. These defenses modify either the model's training or inference procedure to produce minimally perturbed confidence vectors that thwart existing membership inference attacks. We refer to these defenses as ``confidence-masking'' defenses.

\vspace{-6pt}
\paragraph{Outliers in Membership Inference} Most membership inference research is focused on protecting the average-case user's privacy: the success of a membership inference attack is evaluated over a large dataset.
Long \etal~\cite{generalizedmi} focus on understanding the vulnerability of \emph{outliers} to membership inference. They show that some ($<100$) outlier data points can be targeted and have their membership inferred to high (up to $90\%$) precision~\cite{long2017towards,generalizedmi}. 
Recent work explores how overfitting impacts membership leakage from a defender's (white-box) perspective, with complete access to the model~\cite{leino2019stolen}.
\section{Evaluation Setup}\label{ssec:ourthreatmodel}

Because our main goal is to show that label-only attacks can match the success of prior attacks, we consider a similar threat model that matches prior work--except that we restrict the adversary to label-only queries.

As in prior work~\cite{cornellMI}, we assume that the adversary has: (1) full knowledge of the task; (2) knowledge of the target model's architecture and training setup; (3) partial data knowledge, i.e., access to a disjoint partition of data samples from the same distribution as the target model's training data (see below for more details); and (4) knowledge of the targeted points' labels, $y$.

\subsection{Our Threat Model}\label{app:generating-data}
\paragraph{Generating Membership Data}
Some works have explored generating data samples $x$ for which to perform membership inference on, which assumes the least data knowledge~\cite{cornellMI,fredrikson2015model}. These cases work best with minimal numbers of features or binary features because they can take many queries~\cite{cornellMI}. Other works assumes access to the confidence vectors~\cite{fredrikson2015model}. Our work assumes that candidate samples have already been found by the adversary. We leave to future work the efficient discovery of these samples on high-dimensionality data using a label-only query interface.

In our threat model, we always use a disjoint, non-overlapping (\ie no data points are shared) set of samples for training and test data for the target model. The source model uses another two separate subsets of the task's total data pool. Due to the balanced priors we assume, all subsets (\ie the target model training and test sets, and the source model training and test sets) are always of the same size. In the case of CIFAR100, we use the target models training dataset (members) as the source models test dataset (non-members), and vice versa.

\paragraph{Model Architectures}\label{app:model-architectures-and-training}For computer vision tasks, we use two representative model architectures, a standard convolutional neural network (CNN) and a ResNet~\cite{heresnet}.
Our CNN has four convolution layers with ReLU activations. The first two $3\times3$ convolutions have $32$ filters and the second two have 64 filters, with a max-pool in between the two. To compute logits we feed the output through a fully-connected layer with $512$ neurons. This model has $1.2$ million parameters.
Our ResNet-28 is a standard Wide ResNet-28 taken directly from \cite{fixmatch} with $1.4$ million parameters.
On Purchase-100, we use a fully connected neural network with one hidden layer of size $128$ and the $Tanh$ activation function, exactly as in~\cite{cornellMI}. For Texas-100, Adult, and Locations we mimic this model but add a second hidden layer matching the first.

For the attacks from prior work based on confidence vectors, and our new label-only attacks based on data augmentations, we use shallow neural networks as membership predictor models $f$.
Specifically, for augmentations, we use two layers of 10 neurons and LeakyReLU activations~\cite{relu}. The confidence-vector attack models use a single hidden layer of 64 neurons, as originally proposed by Shokri \etal~\cite{cornellMI}. We train a separate prediction model for each class 
We observe minimal changes in attack performance by changing the architecture, or by replacing the predictor model $f$ by a simple thresholding rule. Our combined boundary distance and augmentation attack uses neural networks as well. For simplicity, our decision boundary distance attacks use a single global thresholding rule, $2,500$ queries, and the L2 distance metric. See Section~\ref{ssec:decisionboundarydistance} for more details.

\subsection{On Measuring Success}\label{ssec:on-measuring-success}
Some recent works have questioned the use of (balanced) accuracy as a measure of attack success
and proposed other measures more suited for imbalanced priors: where any data point targeted by the adversary is a-priori unlikely to be a training point~\cite{jayaraman2020revisiting}. 
As our main goal is to study the effect of the model's \emph{query interface} on the ability to perform membership inference, we focus here on the same balanced setting considered in most prior work. 
We also note that the assumption that the adversary has a (near-) balanced prior need not be unrealistic in practice: For example, the adversary might have query access to models from two different medical studies (trained on patients with two different conditions) and might know a-priori that some targeted user participated in one of these studies, without knowing which.

\section{Threat Model}\label{app:threatmodel}

\begin{table*}[t]
    \centering
    \caption{\textbf{Survey of membership inference threat models}.
    $\mathcal{L}$ is the model's loss function, $\tau$ is a calibration term reflecting the difficulty of the sample, $\theta$ are the model parameters centered around $\theta^*$, $\theta_0$ are the parameters on all other datapoints (other than $x$), $\text{aug}(x)$ is a data augmentation of $x$ (e.g., image translation), x' is an adversarial-example of x, and $\dist_h(x, y)$ is the distance from $x$ to the decision boundary.
    Train, data, label, and model knowledge mean, respectively, that the adversary (1) knows the model's architecture and training algorithm, (2) has access to other samples from the training distribution, (3) knows the true label, $y$ for a given $x$, and (4) knows the model parameter values.
    }
    \vspace{0.5em}
    \begin{tabular}{@{} l@{\hskip 1em} c@{\hskip 1em} c@{\hskip 1em} c@{}}
        Query Interface & Attack Feature & Knowledge & Source\\
        \toprule
        confidence vector & $h(x), y$ & train, data, label & \cite{cornellMI} \\
        confidence vector& $h(x)$ & train, data & \cite{long2017towards} \\
        confidence vector & $h(x)$ & -- & \cite{salem2018mlleaks}\\
        confidence vector & $\mathcal{L}\left(h\left(x\right),~y\right)$ & label & \cite{privacyoverfittingmi}\\
        confidence vector & $\mathcal{L}\left(h\left(x\right),~y\right) + \tau(x)$ & label & \cite{sablayrolles2019white} \\
        confidence vector & $-(\theta - \theta_0^*)^T\nabla_\theta\mathcal{L}\left(h\left(x\right),~y\right)$ & train, data, label, model & \cite{sablayrolles2019white} \\
        confidence vector & h(x'), y & train, data, label & \cite{song2019privacy} \\
        label-only & $\argmax h(x), y$  & label & \cite{privacyoverfittingmi}\\\\
        \midrule
        label-only & $\argmax h(\text{aug}(x)), y$ & train, data, label & ours \\
        label-only & $\dist_h(x, y)$ & train, data, label & ours \\
        label-only & $\dist_h(\text{aug}(x), y)$ & train, data, label & ours \\
    \end{tabular}
    \label{tab:priorattacks}
\end{table*}
\vspace{-7pt}

The goal of a membership inference attack is to determine whether or not a candidate data point was used to train a given model. 
In Table~\ref{tab:priorattacks}, we summarize 
different sets of 
assumptions made in prior work about the adversary's knowledge and query access to the model.

\subsection{Adversarial Knowledge}
The membership inference threat model originally introduced by Shokri \etal~\cite{cornellMI}, and used in many subsequent works~\cite{long2017towards,demystify,salem2018mlleaks,song2019privacy,shokriminmax}, assumes that the adversary has \emph{black-box} access to the model $h$ (i.e.,
they can only query the model for its prediction and confidence but not inspect its learned parameters
).
Our work also assumes black-box model access, with the extra restriction
(see Section~\ref{ssec:capabilities} for more details)
that the model only returns (hard) labels to queries.
Though studying membership inference attacks with white-box model access~\cite{leino2019stolen} has merits (e.g., for upper-bounding the membership leakage), 
our label-only restriction inherently presumes a 
black-box 
setting (as otherwise, the adversary could just run $h$ locally to obtain confidence scores). Although we are focused on the label-only domain, our attack methodologies can be applied for analysis in the white-box domain.

Assuming a black-box query interface, there are a number of other dimensions to the adversary's assumed knowledge of the trained model:

\vspace{-6pt}
\paragraph{Task Knowledge} refers to global information about the model's prediction task and, therefore, of its prediction API.
Examples of task knowledge include the total number of classes, the class-labels (dog, cat, etc.), and the input format ($32 \times 32$ RGB or grayscale images, etc.). Task knowledge is always assumed to be known to the adversary, as it is necessary for the classifier service to be useful to a user.

\vspace{-6pt}
\paragraph{Training Knowledge} refers to knowledge about the \textit{model architecture} (e.g., the type of neural network, its number of layers, etc.) and how it was trained (the training algorithm, training dataset size,
etc). 
This information could be publicly available or inferable from a model extraction attack~\cite{tramer2016stealing,Wang2018StealingHI}.

\vspace{-6pt}
\paragraph{Data Knowledge} constitutes knowledge about the 
data that was used to train the target model. 
Full knowledge of the training data renders membership inference trivial because the training members are already known. Partial knowledge may consist in having access to (or the ability to generate) samples from the same or a related data distribution.

\vspace{-6pt}
\paragraph{Label Knowledge} refers to knowledge of the \emph{true} label $y$ for each point $x$ for which the adversary is predicting membership. Whether knowledge of a data point implies knowledge of its true label depends on the application scenario. Salem \etal~\cite{salem2018mlleaks} show that attacks that rely on knowledge of query labels can often be matched by attacks that do not.

\subsection{Query Interface}
\label{ssec:capabilities}

Our paper studies a different query interface than most prior membership inference work.
The choice of query interface ultimately depends on the application needs where the target model is deployed.
We define two types of query interfaces, with different levels of response granularity:

\vspace{-6pt}
\paragraph{Full confidence vectors} 
On a query $x$, the adversary receives the full vector of confidence scores $h(x)$ from the classifier. 
In a multi-class scenario, each value in this vector corresponds to an estimated confidence that this class is the correct label. 
Restricting access to only part of the confidence vector has little effect on the adversary's success~\cite{cornellMI,demystify,salem2018mlleaks}.

\vspace{-6pt}
\paragraph{Label-only} Here, the adversary only obtains the predicted label $y = \argmax_i h(x)_i$, with no confidence scores. This is the minimal piece of information that any query-able machine learning model must provide and is thus the most restrictive query interface for the adversary. Such a query interface is also realistic, as the adversary may only get \emph{indirect} access to a deployed model in many settings. For example, the model may be part of a larger system taking actions based on the model's predictions---the adversary can only observe the system's actions but not the internal model's confidence scores.

In this work, we focus exclusively on the above label-only regime. Thus, in contrast to prior research~\cite{cornellMI,gan,demystify,salem2018mlleaks}, our attacks can be mounted against \emph{any} machine learning service, regardless of the granularity provided by the query interface.

\section{Confidence-Masking Defense Descriptions}\label{app:confidence-masking}
\paragraph{MemGuard}\label{app:memguard} This defense solves a constrained optimization problem to compute a defended confidence-vector $h^{\text{defense}}(x) = h(x) + n$, where $n$ is an adversarial noise vector that satisfies the following constraints: (1) the model still outputs a vector of ``probabilities'', i.e., $h^{\text{defense}}(x) \in [0, 1]^K$ and $\|h^{\text{defense}}(x)\|_1 = 1$; (2) the model's predictions are unchanged, i.e., $\argmax h^{\text{defense}}(x) = \argmax h(x)$; and (3) the noisy confidence vector ``fools'' existing membership inference attacks. To enforce the third constraint, the defender locally creates a membership attack predictor $f$, and then optimizes the noise $n$ to cause $f$ to mis-predict membership.

\paragraph{Prediction Purification}\label{app:prediction-purification} Prediction purification~\cite{yang2020predictionpurification} is a similar defense. It trains a purifier model, $G$, that is applied to the output vector of the target model. That is, on a query $x$, the adversary receives $G(h(x))$. The purifier model $G$ is trained so as to minimize the information content in the confidence vector, whilst preserving model accuracy. While the defense does not guarantee that the model's labels are preserved at all points, the defense is by design incapable of preventing the baseline gap attack, and it is likely that our stronger label-only attacks would similarly be unaffected (intuitively, $G(h(x))$ is just another deterministic classifier, so the membership leakage from a point's distance to the decision boundary should not be expected to change).

\paragraph{Adversarial Regularization}\label{app:adversarial-regularization}
This defense trains the target model in tandem with a defensive membership classifier. This defensive membership classifier is a neural network that accepts both the confidence-vector, $h(x)$, of the target model, and the true label, $y$, that is one-hot encoded. Following the input $h(x)$ there are four fully connected layers of sizes $100$, $1024$, $512$, $64$. Following the input $y$, there are three fully connected layers of sizes $100$, $512$, $64$. The two $64$ neuron layers are concatenated (to make a layer of size $128$), and passed through three more fully connected layers of sizes $256$, $64$, and the output layer of size $1$. ReLU activations are used after every layer except the output, which uses a sigmoid activation.

The defensive membership classifier and the target model are trained in tandem. First the target model is trained a few (here, $3$) epochs. Then for $k$ steps, the defensive membership classifier is trained using an equal batch on members and non-members (which should be different from the held-out set for the target model). After, the target model is trained  on one batch of training data. The target model's loss function is modified to include a regularization term using the output of the defensive classifier on the training data. This regularization term is weighted by $\lambda$.

\section{Description of Common Regularizers}
\label{apx:regularizers}
Dropout~\cite{srivastava2014dropout} is a simple regularization technique, wherein a fraction $\rho \in (0, 1)$ of weights are randomly ``dropped'' (i.e., set to zero) in each training step. Intuitively, dropout samples a new random neural network at each step, thereby preventing groups of weights from overfitting. At test time, the model is deterministic and uses all the learned weights.
We experiment with different dropout probabilities $\rho$.

L1 and L2 regularization simply add an additional term of the form $\lambda \cdot ||w||$ to the model's training loss, where $w$ is a vector containing all of the model's weights, the norm is either L1 or L2, and $\lambda>0$ is a hyper-parameter governing the scale of the regularization relative to the learning objective. Strong regularization (i.e., large $\lambda$) reduces the complexity of the learned model (i.e., it forces the model to learn smaller weights).
We experiment with different regularization constants $\lambda$.

Differential privacy guarantees that any output from a (randomized) algorithm on some dataset $D$, would have also been output with roughly the same probability (up to a multiplicative $e^\epsilon$ factor) if one point in $D$ were arbitrarily modified.
For differential privacy, we use DP-SGD~\cite{dpdeeplearning}, a private version of stochastic gradient descent that clips per-example gradients to an L2 norm of $\tau$, and adds Gaussian noise $\mathcal{N}(0, c^2\tau^2)$ to each batch's gradient. We train target models with fixed parameters $c=0.5$ and $\tau=2$. We train for a varied number of steps, to achieve provable differential privacy guarantees for $10 \leq \epsilon \leq 250$.

\section{Additional Figures}
\label{app:additional-figures}

\begin{figure}[htb]
    \centering
    \includegraphics[width=.95\columnwidth]{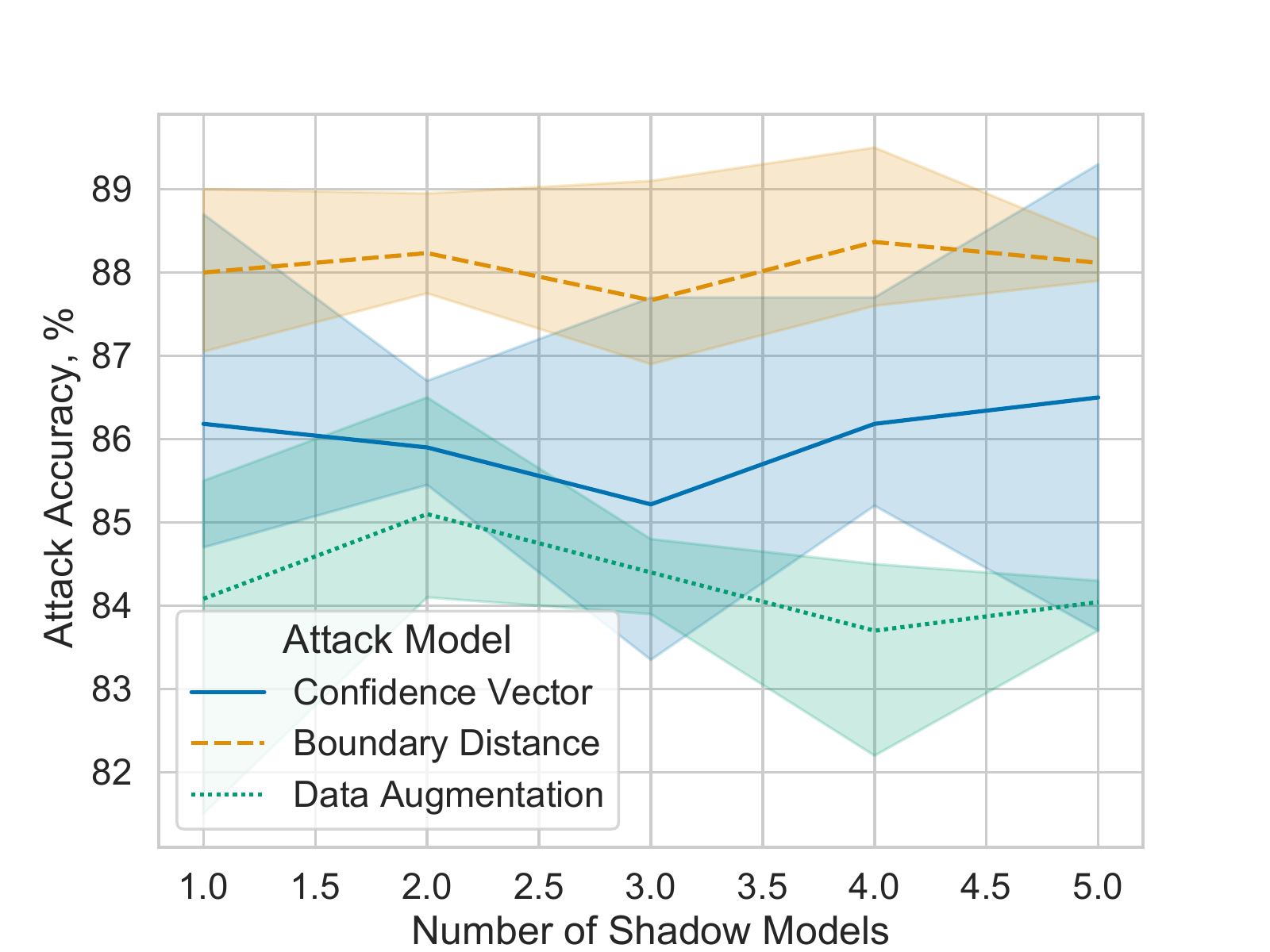}
    \caption{\textbf{Attack accuracy of our label-only attacks for various numbers of shadow models.} Target and source models are trained on 1000 data points from CIFAR-10. The number of shadow models \textbf{does not} have a significant impact on the attack accuracy.
    }
    \vspace{-6pt}
    \label{fig:shadows}
\end{figure}

\begin{figure}[htb]
    \centering
    \includegraphics[width=.95\columnwidth]{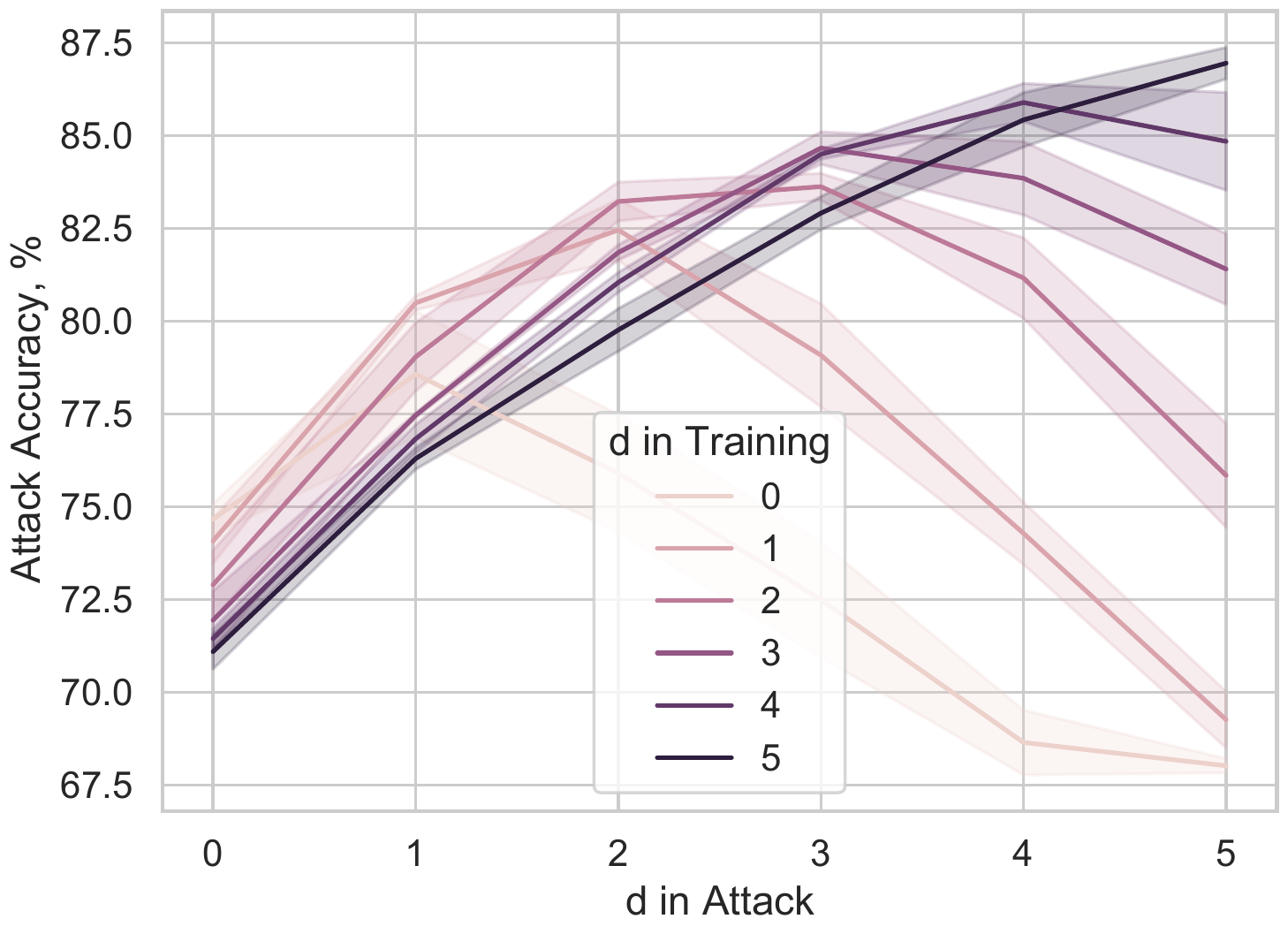}
    \caption{\textbf{Attack accuracy of our translation attack for various choices of $d$.} Target models are trained on 2500 data points from CIFAR-10 with varied sizes of translation augmentations. The attack's accuracy is maximized when it evaluates the same size $d$ of translations as used for training.
    }
    \label{fig:choosed}
\end{figure}

\begin{figure}[htb]
    \centering
    \includegraphics[width=\columnwidth]{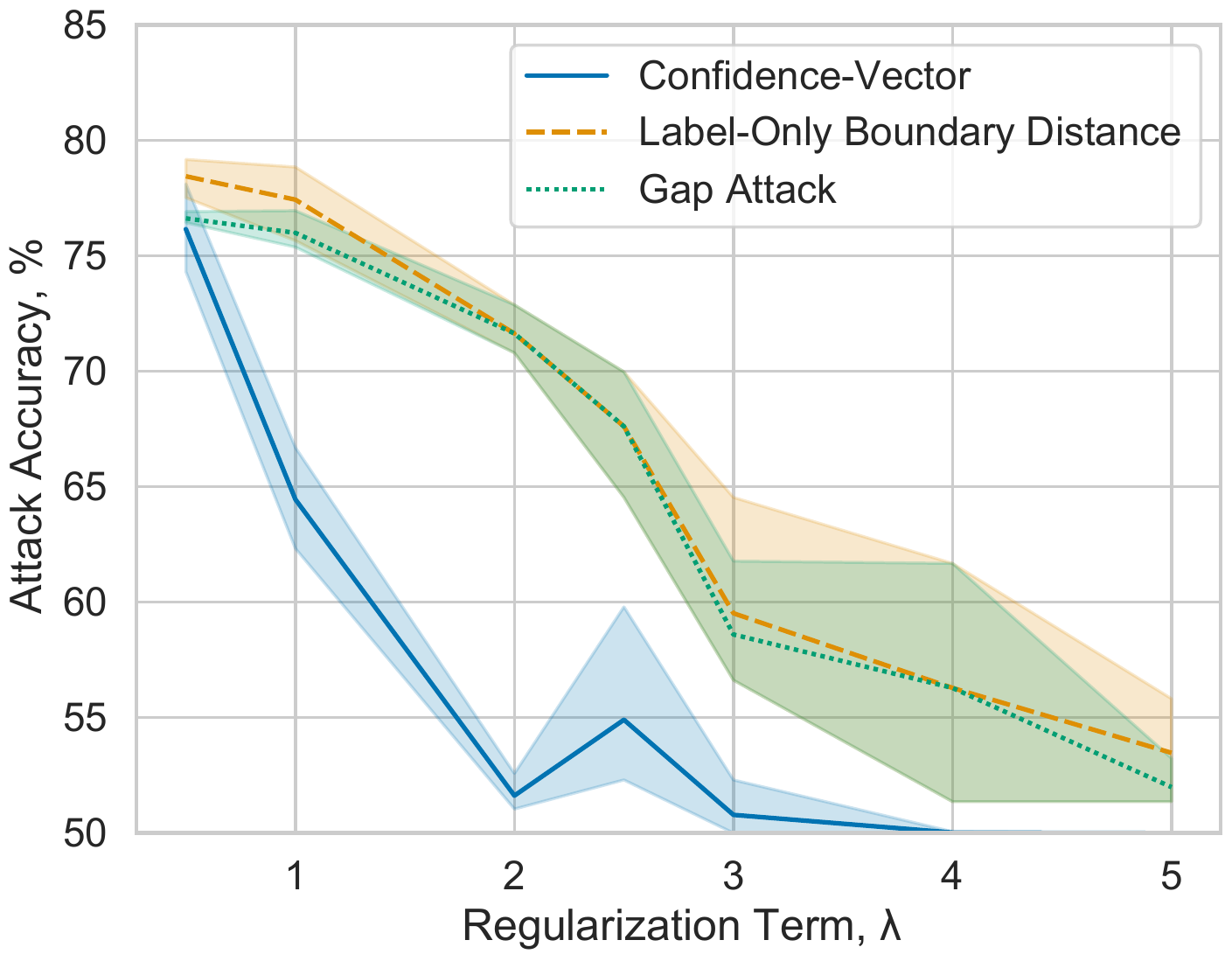}
    \vspace{-2em}
    \caption{\textbf{Accuracy of membership inference attacks on CIFAR-10  models protected with Adversarial Regularization~\cite{adversarialregularization}.} Target models are trained on a subset of 2500 images. We test several values of $k$, the ratio of maximization to minimization steps and find that setting $k=1$ enabled the target model to converge to a defended state. We report results as we vary the second hyper-parameter, $\lambda$, which balances the two training objectives (low training error and low membership leakage). This defense strategy does not explicitly aim to reduce the train-test gap and thus does not protect against label-only attacks. However, we find that this defense prevents attacks from exploiting beyond $3$ percentage points of the gap attack. Test accuracy ranges from $45\%$ to $20\%$, where $\lambda\geq 3$ had a test accuracy below $35\%$.}
    \vspace{-6pt}
    \label{fig:adversarialregularization}
\end{figure}

\begin{figure}[tbh]
    \centering
    {\includegraphics[width=\columnwidth]{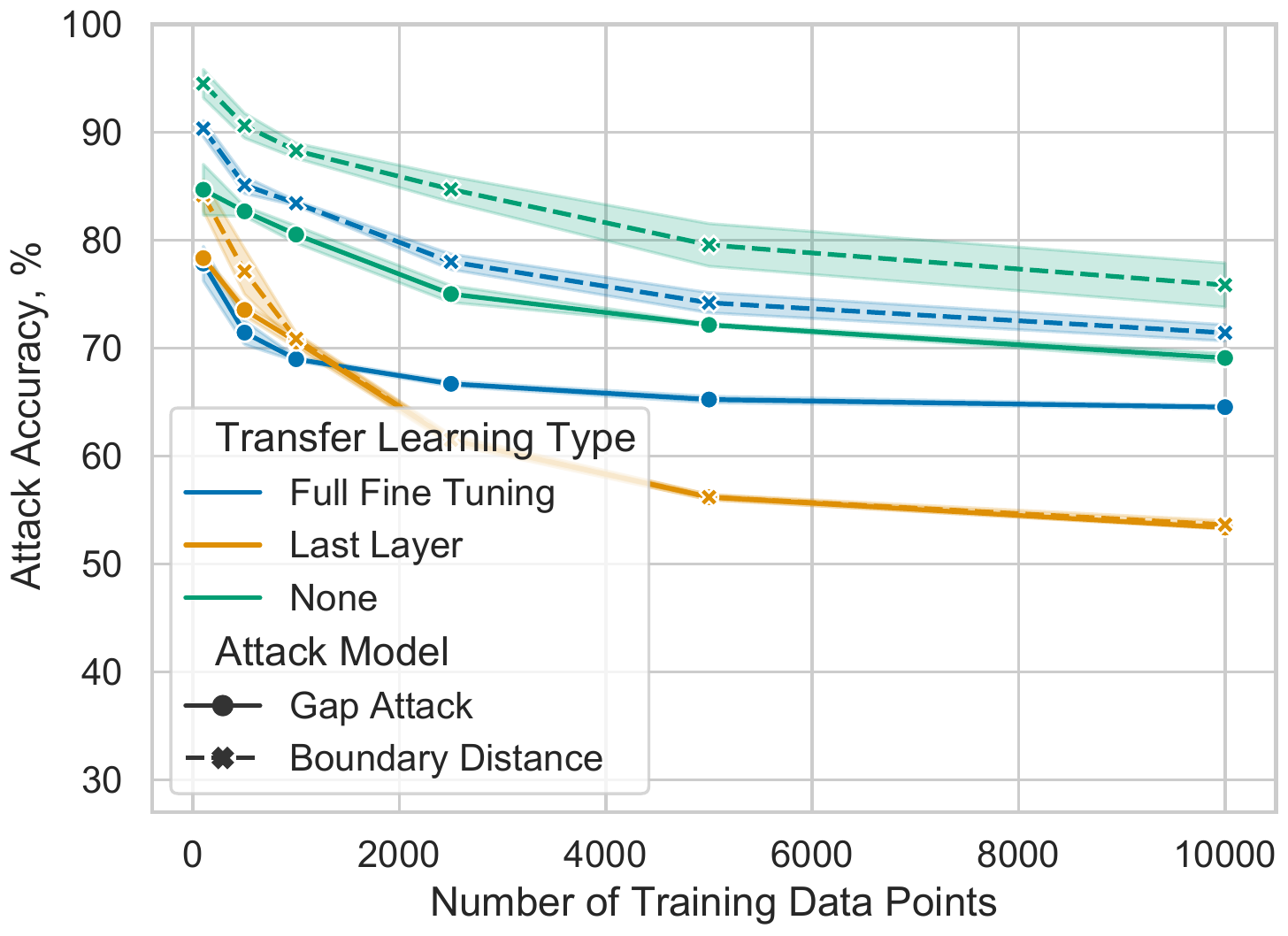}}
    \vspace{-2.5em}
    \caption{\textbf{Accuracy of membership inference attacks on CIFAR-10 models trained with transfer learning.} The source model for transfer learning is trained on all of CIFAR-100. Models are tuned on subsets of CIFAR-10.}
    \label{fig:transferlearn}
\end{figure}

\begin{figure}[tbh]
    \centering
    \vspace{-1em}
    \includegraphics[width=\columnwidth]{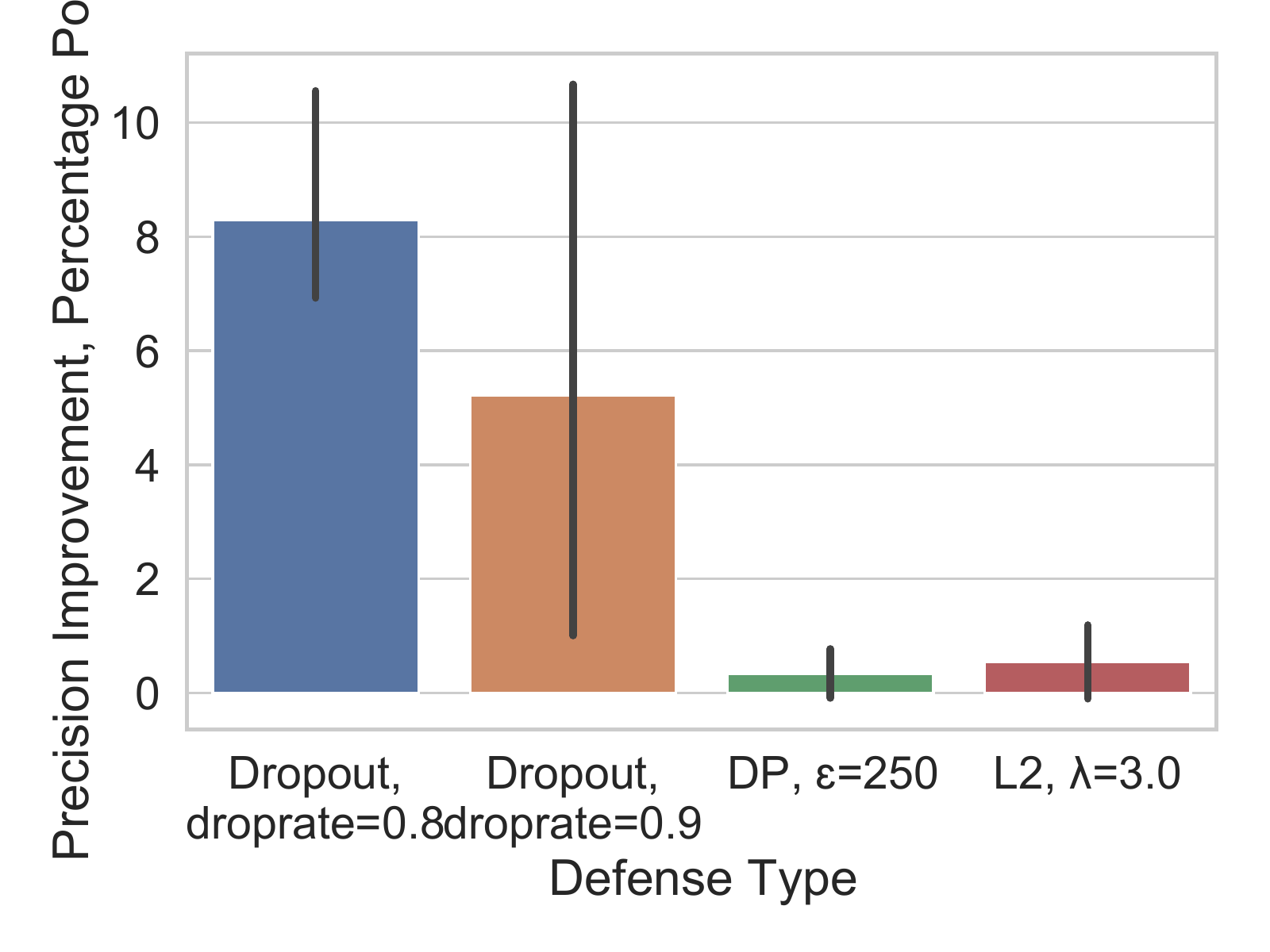}
    \vspace{-3em}
    \caption{\textbf{Outlier membership inference attacks on defended models.} Target and source models are trained on a subset of 2500 points from CIFAR-10. $\beta=2\%$ outliers are identified with less than $\gamma=10$ neighbors. We show precision-improvement from the undefended model, using our label-only boundary distance attack.}
    \label{fig:outliers}
\end{figure}